\documentclass[ALICE,manyauthors]{cernphprep}
\usepackage[comma,square,numbers,sort&compress]{natbib}
\usepackage{hyperref}
\usepackage{lineno}
\usepackage{epsfig} 
\usepackage{subfigure}
\usepackage{verbatim}
\usepackage{color}
\usepackage{soul}
\usepackage{caption}  
\usepackage{array}    
\usepackage{multirow} 
\usepackage[T1]{fontenc}
\usepackage{amsbsy}
\usepackage{here}
\usepackage{romannum}

\newcommand{\sqrts}{\sqrt{s}}

\newcommand{\tev}{\mathrm{TeV}}

\newcommand{\pt}{p_{\mathrm T}}

\begin{document}%

\begin{titlepage}
\PHyear{2022}
\PHnumber{011}      
\PHdate{24 January}  

%

\title{Multiplicity dependence of charged-particle jet production in pp collisions at $\pmb{\sqrt{s}=13}\,\mathbf{TeV}$}
\ShortTitle{Charged-particle jets in pp collisions at $\sqrts=13\,\tev$}

\Collaboration{ALICE Collaboration\thanks{See Appendix~\ref{app:collab} for the list of collaboration members}}
\ShortAuthor{ALICE Collaboration} 

\begin{abstract}
  The multiplicity dependence of jet production in pp collisions at the centre-of-mass energy of $\sqrt{s} = 13\ \mathrm{TeV}$ is studied for the first time. Jets are reconstructed from charged particles  using the anti-$k_\mathrm{T}$ algorithm with resolution parameters $R$ varying from $0.2$ to $0.7$. The jets are measured in the pseudorapidity range $|\eta_{\rm jet}|< 0.9-R$ and in the transverse momentum range $5<p_\mathrm{T,jet}^{\rm ch}<140\ \mathrm{GeV}/c$.
The multiplicity intervals are categorised by the ALICE forward detector V0. 
The $\pt$ differential cross section of charged-particle jets are compared to leading order (LO) and next-to-leading order (NLO) perturbative quantum chromodynamics (pQCD) calculations. It is found that the data are better described by the NLO calculation, although the NLO prediction overestimates the jet cross section below $20\ \mathrm{GeV}/c$.
The cross section ratios for different $R$ are also measured and compared to model calculations. These measurements provide insights into the angular dependence of jet fragmentation.
The jet yield increases with increasing 
self-normalised charged-particle multiplicity. This increase shows only a weak dependence on jet transverse momentum and resolution parameter at the highest multiplicity. 
While such behaviour is qualitatively described by the present version of PYTHIA, quantitative description may require implementing new mechanisms for multi-particle production in hadronic collisions.

\end{abstract}
\end{titlepage}
\setcounter{page}{2}

\section{Introduction}  
\label{sec:intro}
Jets, as sprays of collimated hadrons resulting from fragmentation of high-energy partons
produced in hard scatterings, are multipurpose tools to explore various properties of the strong interaction, via measurements of the strong coupling constant~\cite{Sirunyan:2019crt,
Aaboud:2017alphas}, heavy-flavour production~\cite{Sirunyan:2017ezt}, and the global fit analysis of Parton Distribution Functions (PDF)~\cite{PhysRevD.103.014013}. Jet processes are also pivotal to address fundamental questions such as the validity of factorisation theorems~\cite{Sterman:2014nua,Collins:1989gx,doi:10.1146/annurev.ns.37.120187.002123}, or existence of gluon saturation due to nonlinear QCD dynamics~\cite{Soyez:2006rh}. Jets, produced abundantly at LHC energies due to the large centre-of-mass energy and high luminosity, provide high precision tests of QCD~\cite{Acharya:2018eat,Aaboud:2017wsi,Acharya:2019jyg,Acharya:2019tku}. Moreover, measuring the ``jet quenching'' phenomenon~\cite{Burke:2013yra} observed for jets produced in heavy-ion collisions offers an insight into how high-momentum partons interact with the medium created in the collisions.
Through this interaction one probes QCD at high energy densities and temperatures, where the strongly-interacting matter enters the guark--gluon plasma (QGP) phase.
The observed properties of this matter are consistent with a strongly-coupled, 
low-viscosity fluid of quarks and gluons~\cite{RevModPhys.89.035001,PhysRevLett.105.252302}.

Recent measurements in high-multiplicity pp and pA collisions~\cite{doi:10.1146/annurev-nucl-101916-123209} exhibit several collective effects qualitatively similar to the ones observed in AA collisions for various observables. Such collective behaviour encompasses long-range ($\lvert\Delta\eta\rvert>2$) near-side ($\lvert\Delta\varphi\rvert\approx 0$) two-particle angular correlations, known as the ``ridge''~\cite{Khachatryan:2010gv,PhysRevLett.116.172301,PhysRevC.96.024908,2017193,2013795}; enhanced yield of charged- or identified-particle production in high-multiplicity events with respect to the reference using minimum-bias (MB) charged- or pion-particle production
~\cite{PRC99_2019_024906,ALICE:2017jyt,EPJC80_2020_167,PLB728_2014_25,PLB758_2016_389}; elliptic flow of heavy-flavour hadrons~\cite{Aad:2019aol}; and enhanced baryon production at intermediate transverse momentum ($p_\mathrm{T}\sim 3\ \mathrm{GeV}/c$)~\cite{CMS_ppStrangeness}. 
With the lack of experimental observations of 
jet quenching effect with present accuracy in small collision systems~\cite{Adam:2015hoa,Adam:2016jfp}, the measurements aforementioned raise the difficult but intriguing question of whether these observations arise similarly to heavy-ion collisions, namely from the formation of a hot and dense fluid-like medium, or rather involve other physical mechanisms~\cite{Mace:2018yvl,Mace:2018vwq}. Several theoretical approaches and models have been put forward to explain these QGP-like effects in small systems while accounting for the absence of jet quenching~\cite{JHEP01_2017_140,SJOSTRAND201943,Bierlich:2018xfw,Varga:2018isd,daSilva:2020cyn}, such as multiple parton interactions (MPI)~\cite{Bartalini:2011jp}, string shoving~\cite{Bierlich:2016vgw}, or rope hadronisation~\cite{Bierlich:2014xba}. However, these models cannot explain the measured non-zero elliptic flow at high $p_\mathrm{T}$ from two-particle correlations~\cite{PhysRevC.96.024908}, which is usually attributed to in-medium path-length dependent energy loss~\cite{Shen:flow2017,TYWONIUK201485,Csanad:2016add}.

To deepen our understanding of the mechanisms that are at play in high-multiplicity collisions of small systems, the multiplicity dependence of the charged-jet production has been studied in pp collisions at a centre-of-mass energy $\sqrt{s}=13\ \mathrm{TeV}$ by ALICE. Charged-particle jets are reconstructed from tracks measured at midrapidity using the anti-$k_\mathrm{T}$ clustering algorithm~\cite{A25_RefAntikt} with jet resolution parameters $R$ ranging from $0.2$ to $0.7$. The measured inclusive jet cross sections are compared to model calculations, 
allowing us to test the relative importance of various mechanisms at play in particular hadronisation and MPI to which the lowest transverse momentum jets are most sensitive.

The event activity is quantified by the charged-particle multiplicity measured by the ALICE V0 detector at forward rapidity, in order to minimise the autocorrelations between the event selection and the measured observable.
By taking advantage of the largest integrated luminosity collected so far by the ALICE experiment, this work extends previous measurements~\cite{Acharya:2019tku,ALICE_chJets7TeV,Acharya:2018eat} to higher collision energy, broader jet kinematic range, larger jet radii (up to $R = 0.7$), and higher event activities.
A complementary insight is thereby provided to similar CMS~\cite{Chatrchyan:2013ala} and ALICE measurements focusing on the soft sector based on transverse spherocity~\cite{EPJC79_2019_857} and two-particle correlations~\cite{Abelev:2014mva}. 
Thanks to these new measurements, stronger constraints are placed on models that describe fundamental mechanisms of particle production in hadronic collisions.

The results presented in this article test
whether the seemingly universal pattern of the self-normalised production of hard probes as a function of event activity, emerging from the ALICE measurements of $\mathrm{J/\psi}$~\cite{Abelev:2012rz,Acharya:2020pit,Acharya:2020giw}, $\rm D$~\cite{Adam:2016mkz}, and $\rm B$-meson production~\cite{Adam:2015ota}, holds also 
 for jets. 
 While 
 this self-normalised hard probe production pattern could be ascribed to mere collision geometry up to a certain multiplicity, a new regime of high-density gluon configurations is expected to set in at the highest multiplicities~\cite{AZARKIN2014244}. 

The article is organised as follows: Sec.~\ref{sec:detector} describes the ALICE detectors, the experimental data samples, and Monte-Carlo simulations used for this analysis; Sec.~\ref{sec:jetrec} discusses the multiplicity selection and the jet reconstruction methods; Sec.~\ref{sec:Syst} outlines the unfolding corrections and the estimation of the systematic uncertainties; Sec.~\ref{sec:results} presents the results; and finally, Sec.~\ref{sec:conclusion} gives a summary and outlook.

\section{Experimental setup and data sample}  
\label{sec:detector}

ALICE is the dedicated heavy-ion experiment at the LHC. A detailed description of the ALICE apparatus can be found in Ref.~\cite{ALICEexp}. In the following, only the detector components used in the data analysis presented in this article are described.

The ALICE apparatus comprises a central barrel (pseudorapidity coverage $|\eta|<0.9$ over the full azimuth) situated in a uniform $0.5\ \mathrm{T}$ magnetic field along the beam axis ($z$), which is supplied by the large L3 solenoid magnet~\cite{Wittgenstein:1989rw}. The central barrel 
contains a set of tracking detectors: a six-layer silicon Inner Tracking System (ITS) surrounding the beam pipe, and a large-volume ($5\ \mathrm{m}$ length, $0.85\ \mathrm{m}$ inner radius and $2.8\ \mathrm{m}$ outer radius) cylindrical Time Projection Chamber (TPC). The first two layers of the ITS are instrumented with high-granularity Silicon Pixel Detectors (SPD), followed by two layers composed of Silicon Drift Detectors (SDD), and finally, the two outer layers are made of double-sided Silicon micro-Strip Detectors (SSD). 
In the forward and backward rapidity regions, a pair of plastic scintillator counters called V0A and V0C are positioned on each side of the interaction point, covering pseudorapidity ranges ${2.8<\eta<5.1}$ and ${-3.7<\eta<-1.7}$, respectively. The V0 system provides the interaction trigger for the whole experiment, and is further used to suppress machine-induced background events~\cite{ALICEV0}.

The measurement was based on the data from pp collisions at a centre-of-mass energy of $\sqrt{s}=13~\,\mathrm{TeV}$ collected between 2016 and 2018. During this period, MB events were selected online using the high purity V0-based MB trigger~\cite{ALICEperf}, 
which required a charged-particle signal coincidence in the V0A and V0C arrays. The visible cross section satisfying the MB trigger requirement was determined in a van der Meer scan~\cite{BalaguravdM,vanderMeer}.
The integrated luminosity of the used sample, measured with V0, is $8.12 \pm 0.16~\rm nb^{-1}$ for 2016, $10.67 \pm 0.29~\rm nb^{-1}$ for 2017, and $13.14 \pm 0.27~\rm nb^{-1}$ for 2018, respectively. The luminosity uncertainty was evaluated to be 1.6~\% by taking into account the correlations during the combination of the samples ~\cite{ALICE-PUBLIC-2021-005}.
For the offline analysis, further event selection was made by requiring a primary vertex position within $\pm 10\ \mathrm{cm}$ in the longitudinal direction around the nominal interaction point to ensure full geometrical acceptance in the ITS for ${|\eta|<0.9}$. Pile up, i.e. the average number of simultaneous interactions per bunch crossing, was maintained well below unity through beam separation in the horizontal plane. Residual pile-up events were rejected based on a multiple vertex finding algorithm using tracking information from the SPD.
The final corresponding data sample consists of $2.2\times 10^9$ events after the trigger and offline selection.

Reconstructed tracks with transverse momenta larger than $0.15\ \mathrm{GeV}/c$ were selected in the analysis. The track selection criteria was optimised for good momentum resolution and minimal contamination from secondary particles, as described in Refs.~\cite{Acharya:2019tku,Acharya:2019jyg}.
To ensure a uniform $(\eta,\varphi)$ distribution in the regions where the SPD was inactive, tracks with no hit in either of the two SPD layers were constrained to the primary vertex. 
The tracking efficiency estimated from a full detector simulation amounts to $80\%$ for $p_\mathrm{T}>0.4\ \mathrm{GeV}/c$, decreasing to $60\%$ at $0.15\ \mathrm{GeV}/c$. The transverse momentum resolution is better than $3\%$ for tracks with $\pt$ below $1\ \mathrm{GeV}/c$, and increases linearly up to $10\%$ at $p_\mathrm{T}=100\ \mathrm{GeV}/c$.

The response of the ALICE detector to produced particles was evaluated using GEANT3~\cite{A32_RefGeant3}. 
Based on this response, measured distributions were corrected for instrumental effects, see Sec.~\ref{subsec:corr}.
In this paper, the default Monte Carlo (MC) event generator used for comparison with measurements is the PYTHIA~8.125 general-purpose Leading Order (LO) MC generator (hereafter referred to as PYTHIA8) with the Monash-2013 set of tuned parameters (tune)~\cite{Pythia8Monash} for the underlying event (UE) and NNPDF2.3 LO PDF set~\cite{PDF_NNPDF},
while MPI and Colour Reconnection (CR) models being enabled. Furthermore, in order to reduce the large theoretical uncertainties affecting the computations at LO in perturbative QCD, like the residual dependence of the unphysical factorisation and renormalisation scales, jet production at Next-to-Leading Order (NLO) accuracy was obtained within the POWHEG framework~\cite{PDF_CT14NLO}.
Unlike pure fixed order calculations, POWHEG interfaces NLO calculations with PYTHIA8 parton showers to generate exclusive final states. The particle level outputs of such simulations were then directly compared with the experimental data, which were corrected for instrumental effects.

\section{Multiplicity selection and jet reconstruction}
\label{sec:jetrec}
\subsection{Multiplicity selection} 
\label{subsec:mult}
\begin{figure*}[htbp!]
 \begin{center}
 \includegraphics[width=0.9\textwidth]{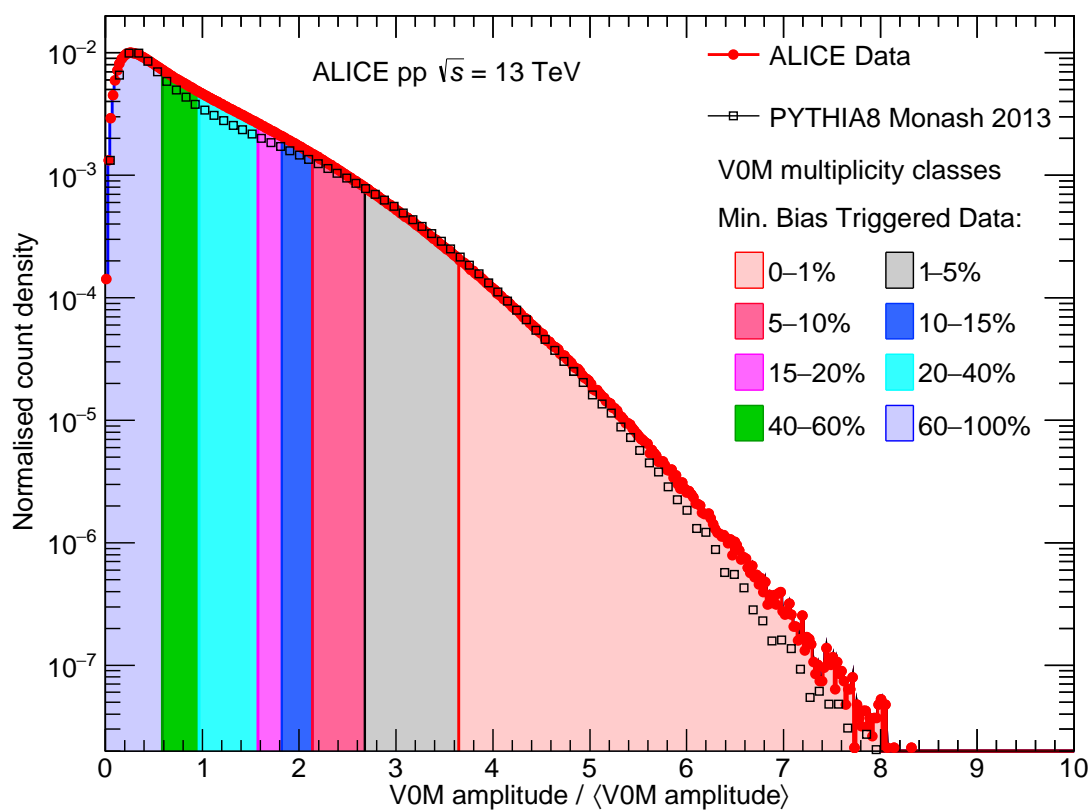}
 \end{center}
 \caption{Scaled V0M distribution which is used to determine the forward multiplicity classes in pp collisions at $\sqrt{s}$ = 13 TeV. The colour shaded areas represent V0M multiplicity classes obtained from real data. The PYTHIA8 distribution is shown with the open black markers.  }
 \label{fig:V0Mmultiplicity}
\end{figure*}

In order to study the multiplicity dependence of inclusive charged-particle jet production, the
MB sample 
was divided into event classes based on the ``V0M amplitude'' that is proportional to the total number of charged particles passing through the V0A and V0C detectors. 
The distribution of the self-normalised V0M amplitude from data and the PYTHIA8 event generator is shown in Fig.~\ref{fig:V0Mmultiplicity}. The distribution is normalised to its average value, $\left \langle\rm V0M~amplitude\right \rangle$, to reduce the sensitivity of the multiplicity percentile determination on the amplitude. PYTHIA8 MC does not reproduce the measured multiplicity distribution, as was already reported in Ref.~\cite{Acharya:2020kyh}. To reduce the potential model-dependent bias, corrections of the multiplicity dependent jet yields were done using a data-driven method instead of pure MC samples, which is discussed in Sec.~\ref{subsec:corr}. 

The event classes used in the analysis and the corresponding midrapidity charged-particle densities for experimental data are summarised in Table~\ref{tab:V0Mmultiplicity}.
The multiplicity classes were defined in terms of percentile intervals of experimental V0M amplitude $/\left \langle\mathrm{V0M~amplitude}\right \rangle$ as shown in Fig.~\ref{fig:V0Mmultiplicity}. The average charged-particle multiplicity densities in MB 
pp collisions and for events of a given multiplicity class were obtained by integrating the corresponding fully corrected $\pt$\ spectra given in Ref.~\cite{Acharya:2020kyh}.
When comparing the data to MC predictions, the multiplicity percentile was calculated from data and MC using their respective self-normalised distribution accordingly in order to minimise the difference observed in the V0M amplitude distribution. 
The 0 -- 1\% range corresponds to the highest multiplicity class (\Romannum{1}), while the 60 -- 100\% interval corresponds to the lowest multiplicity class (\Romannum{8}).

\begin{table}[bhtp!]  
\centering  
\caption{Average charged-particle pseudorapidity densities at midrapidity $\left \langle \mathrm{d}N_{\rm ch}/\mathrm{d}\eta \right \rangle$ from data for inclusive events and different V0M multiplicity classes~\cite{Acharya:2020kyh}.}  

\begin{tabular}[t]{l|l|c|c}

\hline\hline
Class    & {V0M percentile} & {V0M amplitude $/\left \langle \mathrm{V0M~amplitude}\right \rangle$}       &   {$\left \langle \mathrm{d}N_{\rm ch}/\mathrm{d}\eta \right \rangle_{\left |\eta \right |<1}$} \\

\hline 
 MB  &  0--100\%    &                   &  $6.93$ $\pm$ $0.09$ \\

\Romannum{1}      & 0--1\%  & $\geq 3.66$      & $26.01$ $\pm$ $0.34$ \\

\Romannum{2}      & 1--5\%  &  2.68--3.66         & $19.99$ $\pm$ $0.24$ \\

\Romannum{3}      & 5--10\%  &  2.15--2.68          & $16.18$ $\pm$ $0.20$ \\

\Romannum{4}      &10--15\%  &  1.84--2.15          & $13.78$ $\pm$ $0.18$ \\

\Romannum{5}      &15--20\%  &  1.59--1.84          & $12.01$ $\pm$ $0.16$ \\

\Romannum{6}      &20--40\%  &  0.97--1.59          &  $9.18$ $\pm$ $0.10$ \\

\Romannum{7}      &40--60\%  &  0.59--0.97          &  $5.78$ $\pm$ $0.06$ \\

\Romannum{8}      &60--100\%  &  0--0.59          &  $2.94$ $\pm$ $0.03$ \\

\hline\hline
\end{tabular}
\label{tab:V0Mmultiplicity} 
\end{table} 

\subsection{Jet reconstruction} 
\label{subsec:reco}

Jets were reconstructed from tracks with $p_{\rm T,track} > 0.15\ \mathrm{GeV}/c$ and $\left |\eta_{\rm track} \right |<0.9$, using the anti-$k_{\rm T}$ sequential recombination algorithm~\cite{A25_RefAntikt} from the FastJet package~\cite{ref:fastjet}. 
The jet transverse momenta were calculated using boost-invariant $p_{\rm T}$ recombination scheme as a scalar sum of their constituent transverse momenta. Jet resolution parameters were varied in the range from $R=0.2$ to $0.7$. The pseudorapidity of the reconstructed jets was limited to  
$\left |\eta_{\rm jet} \right |<0.9 - R$ to ensure they remain in the fiducial acceptance~\cite{Aad:2011gn}. The transverse momentum range of the inclusive charged-particle jet spectra spans from $5$ to $140\ \mathrm{GeV}/c$. The spectra measured in V0M multiplicity classes have the upper limit $100\ \mathrm{GeV}/c$. 
The cross section of inclusive jet production was measured as a function of $\pt$ considering the van der Meer minimum bias visible cross section mentioned in Sec.~\ref{sec:detector}.
For the multiplicity dependence study, the per-event jet yield was measured as a function of multiplicity classes defined by the V0M percentile intervals. 
In addition, the integrated jet yield was calculated as a function of the charged-particle density $\mathrm{d}N_{\rm ch}/\mathrm{d}\eta$, self-normalised similarly to other ALICE measurements~\cite{Abelev:2012rz,Acharya:2020pit,Acharya:2020giw,Adam:2016mkz,Adam:2015ota}.    

The measured jets are inevitably affected by the UE activity originating from MPI, fragmentation of beam remnants, as well as initial- and final-state radiation~\cite{ALICE:2011ac}. In pp collisions, the UE effect on jet measurements is rather limited~\cite{Acharya:2019tku}. However, since the UE contribution depends on event multiplicity, the measured jets were not affected in the same way for events falling in different multiplicity classes. 
In order to perform fair comparisons between different multiplicity intervals, the results presented in this paper include the UE subtraction. 

The UE contribution to the charged-particle jet $\pt$ was estimated event-by-event using the same approach as in previous measurements in pp~\cite{UErhoCMS7tev} and p--Pb collisions~\cite{Adam:2015hoa,Adam:2016jfp}. 
The background density $\rho_{\rm ch}$ was determined using the $k_{\rm T}$ algorithm~\cite{A27_RefFastjet} with a fixed radius of 0.2, taking into account only jets containing at least one physical particle, while removing the two $k_{\rm T}$ 
clusters of highest transverse momentum to limit the impact of the jet signal on the underlying event estimation. 
The background density $\rho_{\rm ch}$ is then used to subtract the average background from each jet in the same event: 
$p_{\rm T}^{\rm corr} = p_{\rm T}^{\rm raw} - \rho_{\rm ch} \cdot A$, where $A$ is the jet area. 

\begin{figure*}[htbp!]
 \begin{center}
  \includegraphics[width=1.0\textwidth]{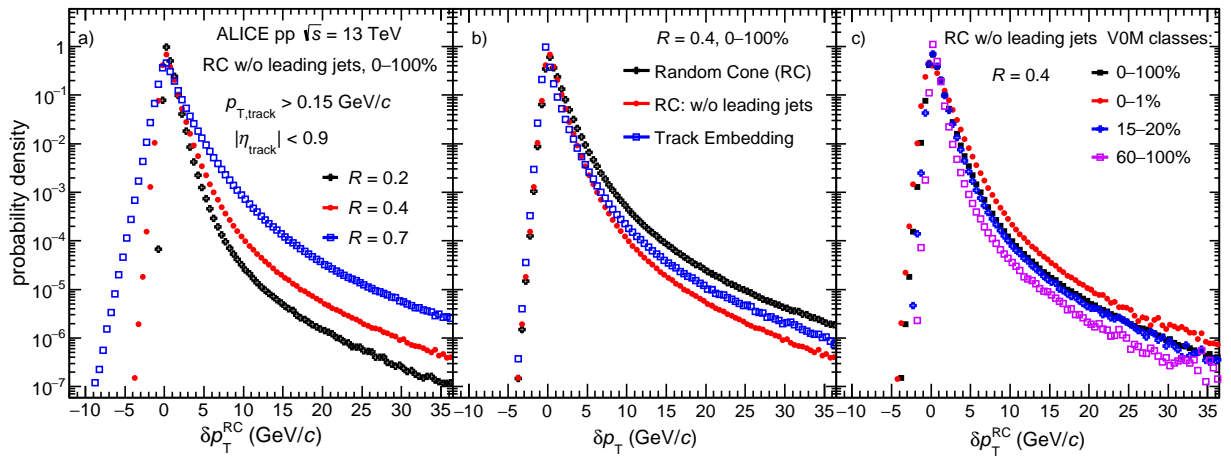}
 \end{center}
 \caption{a) Comparison of the $\delta p_{\rm T}$ distribution obtained for different random cone radii ($R=0.2, 0.4, 0.7$). b) Comparison of the $\delta p_{\rm T}$ distribution with the RC method (including and excluding two leading jets) and the track embedding approach for $R=0.4$. c) Comparison of measured $\delta p_{\rm T}$ distribution using RC method without
 leading jets for $R=0.4$ in different multiplicity classes .}
 \label{fig:deltapt}
\end{figure*}

During the subtraction of the average UE background of each jet, the local background fluctuations smear the reconstructed jet transverse momentum.
To study jet-by-jet fluctuations of the background, the Random Cone (RC) method was used~\cite{ref:pbpb276}. In this method, cones of radius $R$ positioned at random ($\eta, \varphi$) coordinates within the detector acceptance (fiducial region) were generated in each event. The sum of the charged-particle $p_{\rm T}$ in a given cone was then compared to the expected average background obtained from $\rho_{\rm ch}$ as follows: 
\begin{equation}
    \delta p_{\rm T}^{\rm RC} = \sum ^{\rm RC}p_{\rm T,track}-\rho_{\rm ch}\pi R^{2},
    \label{delta-pt-equation}
\end{equation}
where the sum runs over track $p_{\rm T}$ inside the random cone, and $\pi R^{2}$ is the area of the random cone.
The width of $\delta p_{\rm T}^{\rm RC}$ 
is a measure of the momentum smearing due to local background fluctuations~\cite{Bkg_ALICE}. To minimise the influence of signal jets on the $\delta p_{\rm T}^{\rm RC}$ distribution, a minimum distance from the random cone to the two highest momentum jets (leading jets) in the event was required. The $\delta p_{\rm T}^{\rm RC}$ distribution, obtained for different cone radii $R$ in events with excluded leading jets, are shown in Fig.~\ref{fig:deltapt} a).
It clearly shows stronger background fluctuations with increasing jet radius, as expected due to the larger number of particles within the jet cone.

An alternative embedding method was used to quantify the background fluctuations. In this procedure, a probe track was embedded into an event~\cite{ALICE:2021wct}. The azimuthal angle of the probe track was required to be perpendicular to the jet ($\varphi_{\rm track}^{\rm emb} = \varphi_{\rm ch\,jet} + \pi/2$) while retaining its $\eta$ value ($\eta_{\rm track}^{\rm emb} = \eta_{\rm ch\,jet}$). The transverse momentum of the probe track ($p_{\rm T}^{\rm emb}$) was uniformly chosen between $0$ and $100\ \mathrm{GeV}/c$. After embedding the probe track into the event, the jet finding algorithm was relaunched with the same background subtraction method as described above. 
The embedded $\delta p_{\rm T}^{\rm emb}$ was evaluated in a similar way to Eq.~\ref{delta-pt-equation} after removing the momentum of the embedded probe track:

\begin{equation}
    \delta p_{\rm T}^{\rm emb} = p_{\rm T,ch\,jet}^{\rm raw,emb}-\rho_{\rm ch}A_{\rm ch\,jet}^{\rm emb} -p _{\rm T}^{\rm emb},
    \label{delta-pt-probe-equation}
\end{equation}
where $p_{\rm T,ch\,jet}^{\rm raw,emb}$ and $A_{\rm ch\,jet}^{\rm emb}$ are the transverse momentum and area of the reconstructed jet with the embedded probe track. 

The background fluctuations determined by the random cone and the track embedding method are presented in Fig.~\ref{fig:deltapt} b). While the distributions obtained from the different methods show very similar negative tails, the tail on the right-hand side of the distribution caused by real jets is much less pronounced when a minimum distance to two leading jets is required. Therefore, the $\delta p_{\rm T}$ distribution from the RC method without two leading jets was used as default to build up the background fluctuation response, while the track embedding method was used for the assessment of the systematic uncertainty on the background fluctuation estimate. Figure~\ref{fig:deltapt} c) compares the $\delta p_{\rm T}$ distribution in different multiplicity intervals with the RC method when avoiding that the cone overlaps with the two leading jets. The figure suggests that the magnitude of local background fluctuations grows with imposed multiplicity bias, as expected.

\section{Corrections and systematic uncertainties} 
\label{sec:Syst}
The differential charged-particle jet cross section after UE subtraction was corrected for jet energy scale smearing due to local background fluctuations and detector effects. 
The resulting cross section unfolded to charged particle level allows for a direct comparison with the theoretical predictions. 
These unfolding correction approaches are explained in Sec.~\ref{subsec:corr}.
The corresponding systematic uncertainties are presented in Sec.~\ref{subsec:sys}.

\subsection{Unfolding corrections} 
\label{subsec:corr}
The measured jet momentum is affected by background fluctuations and a variety of instrumental effects, including finite momentum resolution and track reconstruction efficiency. The jet spectrum was corrected for these effects to obtain the particle-level jet spectrum through an unfolding procedure~\cite{ref:pbpb276}. 
First, the particle-level true jets were constructed from the PYTHIA8 (Monash 2013) event generator~\cite{Pythia8Monash}, by selecting only those stable charged particles defined as all particles with a mean proper lifetime larger than $1\ \mathrm{cm}/c$, and excluding the decay products of these particles~\cite{ALICE_stable_particle}. 
Next, jets were reconstructed at detector level from tracks coming from MC particles propagated through the GEANT3 model of the ALICE setup. The corresponding jet energy scale residual and resolution were estimated to be about $20$\% for jet transverse momentum larger than $10~\mathrm{GeV}/c$ with jet resolution parameter $R=0.4$.

Based on the geometrical matching between the corresponding jets at the particle level and detector level, the detector level response matrix was constructed. The response matrix employed in the unfolding procedure was a combination of the detector response matrix and the background fluctuation response matrix. The background fluctuation response matrix utilises the $\delta p_{\rm T}$ distribution obtained from data, based on distributions shown in Fig.~\ref{fig:deltapt}. This data-driven approach ensures that the response matrix used for unfolding reflects the accurate multiplicity dependence. Finally, the measured jet spectrum was unfolded using the combined response matrix which corrects for background fluctuations as well as detector effects.

In this analysis, the default unfolding method was based on the Singular Value Decomposition (SVD) approach as implemented in the RooUnfold package~\cite{Adye:2011gm}. The regularisation parameter $k$, which suppresses high-frequency variations in the unfolded result, was selected by examining the $d$-vector distribution~\cite{A44_unfold-svd}. In addition to the SVD unfolding, Bayesian unfolding~\cite{DAgostini:1994zf} was also used for systematic uncertainty evaluation. Consistent results were obtained between both unfolding methods. 
To validate the unfolding process and identify potential biases, closure tests that compare 
the unfolded distributions to the particle-level true distributions were performed. The consistency of the unfolding procedure was also checked by folding the solution to detector level and comparing it to the measured raw spectrum. In both cases, no significant difference was found.

\subsection{Systematic uncertainties} 
\label{subsec:sys}
The sources of systematic uncertainty that affect the jet measurement were divided into the following categories: tracking efficiency and $p_{\rm T}$ resolution, unfolding corrections, background fluctuations, contamination from secondary particles and normalisation. Additionally, in the case of multiplicity dependent measurements, a systematic uncertainty due to multiplicity estimation was also considered. 
All uncertainties from these sources were considered as uncorrelated, except the uncertainties of the unfolding corrections that come from several correlated sources. 
Therefore, the uncertainty of the unfolding category is calculated by varying each source and calculating the RMS of all variations.
Then, all systematic uncertainty categories were treated separately and their respective contributions we added in quadrature. 

Table~\ref{tab:inclusivesys} summarises the contributions to the systematic uncertainties for the inclusive jet cross section in a few selected jet $p_{\rm T}$ and $R$ bins. Table~\ref{tab:multsys} summarises the systematic uncertainties for multiplicity-dependent jet production in a few selected jet $p_{\rm T}$ and multiplicity intervals for a jet reconstruction resolution parameter of $R = 0.4$.

The total systematic uncertainty increases with jet $p_{\rm T}$ and $R$, and
its evolution is similar for all jet radii and multiplicity intervals that are not listed here. 
In the following, the individual sources of systematic uncertainties listed in Table~\ref{tab:inclusivesys} and Table~\ref{tab:multsys} are briefly described. 

\begin{table}[htbp!]
\caption{Summary of systematic uncertainties on inclusive jet cross section for three selected jet transverse momentum bins with different resolution parameters.}
\centering  
\resizebox{\columnwidth}{!}{
\begin{tabular}{p{0.12\textwidth}<{\centering}|p{0.1\textwidth}<{\centering}p{0.1\textwidth}<{\centering}p{0.1\textwidth}<{\centering}p{0.1\textwidth}<{\centering}p{0.11\textwidth}<{\centering}p{0.13\textwidth}<{\centering}p{0.11\textwidth}<{\centering}p{0.1\textwidth}<{\centering}}
\hline
Resolution parameter &  Jet $p_\mathrm{T}$ bin (GeV/$c$) & Tracking efficiency (\%) & Track $p_{\rm T}$ resolution (\%) & Unfolding (\%) & Background fluctuation (\%) & Normalisation (\%) & Secondaries (\%) & Total (\%) \\ 
\hline 
\multirow {3}{*}{   $R = 0.2$   }
&  5--6   &  2.6 & 0.2 & 0.3 & 2.1 & 1.6 & 2.2 & 4.5     \\ \cline{2-9}
&  30--40 &  5.7 & 0.3 & 1.7 & 0.3 & 1.6 & 2.2 & 6.6     \\ \cline{2-9}
& 85--100 &  6.9 & 1.0 & 2.6 & 0.2 & 1.6 & 2.5 & 8.1     \\ \cline{2-9} 
\hline 
\multirow {3}{*}{   $R = 0.4$  }
&  5--6   &  4.2 & 0.4 & 0.2 & 5.2 & 1.6 & 1.7 & 7.2     \\ \cline{2-9}
&  30--40 &  7.4 & 0.4 & 3.4 & 2.5 & 1.6 & 2.3 & 9.1    \\ \cline{2-9}
& 85--100 &  8.2 & 0.7 & 4.5 & 1.1 & 1.6 & 2.6 & 10.0    \\ \cline{2-9} 
\hline
\multirow {3}{*}{   $R = 0.7$   }
&  5--6   &   3.9 & 0.7 & 0.5 & 3.1 & 1.6 & 2.7 & 6.1    \\ \cline{2-9}
&  30--40 &   9.8 & 1.0 & 3.7 & 1.1 & 1.6 & 2.8 & 11.2   \\ \cline{2-9}
& 85--100 &   10.6 & 1.4 & 5.8 & 0.4 & 1.6 & 2.9 & 12.7  \\ \cline{2-9} 
\hline 
\end{tabular}
}
\label{tab:inclusivesys} 
\end{table} 
\begin{table}[htbp!]
\caption{Summary of systematic uncertainties on jet production for three jet transverse momentum bins with different multiplicity percentile intervals for jets with resolution parameter $R = 0.4$.}
\centering  
\resizebox{\columnwidth}{!}{
\begin{tabular}{p{0.12\textwidth}<{\centering}|p{0.1\textwidth}<{\centering}p{0.1\textwidth}<{\centering}p{0.1\textwidth}<{\centering}p{0.1\textwidth}<{\centering}p{0.11\textwidth}<{\centering}p{0.15\textwidth}<{\centering}p{0.11\textwidth}<{\centering}p{0.1\textwidth}<{\centering}}
\hline
Multiplicity percentile &  Jet $p_\mathrm{T}$ bin (GeV/$c$) & Tracking efficiency (\%) & Track $p_{\rm T}$ resolution (\%)& Unfolding (\%) & Background fluctuation (\%) & Multiplicity determination (\%) & Secondaries (\%) & Total (\%) \\ 
\hline 
\multirow {3}{*}{ \shortstack{ 0--1\%} }
&  5--6   &  3.9 & 0.1 & 2.3 & 5.2 & 5.7 & 1.4 & 9.1      \\ \cline{2-9}
&  30--40 &  7.4 & 1.7 & 3.0 & 2.1 & 1.7 & 2.2 & 8.8      \\ \cline{2-9}
& 85--100 &  8.2 & 1.9 & 6.0 & 0.8 & 0.3 & 2.5 & 10.7     \\ \cline{2-9}  
\hline 
\multirow {3}{*}{ \shortstack{ 10--15\%} }
&  5--6   &  4.2 & 0.1 & 1.0 & 5.2 & 5.5 & 1.8 & 8.9      \\ \cline{2-9}
&  30--40 &  7.4 & 1.6 & 4.0 & 1.9 & 0.8 & 2.2 & 9.1      \\ \cline{2-9}
& 85--100 &  8.2 & 1.6 & 5.2 & 0.9 & 0.3 & 2.5 & 10.2     \\ \cline{2-9} 
\hline 
\multirow {3}{*}{  \shortstack{ 60--100\%} }
&  5--6   &   4.5 & 0.1 & 1.0 & 3.2 & 1.9 & 2.0 & 6.3     \\ \cline{2-9}
&  30--40 &   7.6 & 1.3 & 3.6 & 0.6 & 1.6 & 2.4 & 9.0     \\ \cline{2-9}
& 85--100 &   8.2 & 1.3 & 4.9 & 0.2 & 0.4 & 2.6 & 10.0    \\ \cline{2-9} 
\hline
\end{tabular}
}
\label{tab:multsys} 
\end{table} 

The dominant source of systematic uncertainty comes from track reconstruction efficiency since it directly affects the jet energy scale and resolution.
The systematic uncertainty on tracking efficiency was estimated to be 3\% based on variations of the criteria used in the track selection~\cite{Acharya:2019tku}. To evaluate the effect of this uncertainty on the measured jet spectra, a new detector response matrix was computed by generating a PYTHIA simulation that accounts for a 3\% reduction of tracking efficiency, and then used in the unfolding procedure. The difference between the spectra corrected with the default response matrix and with the response matrix obtained with the decreased tracking efficiency was taken as a systematic uncertainty. The relative uncertainty on the jet spectra caused by tracking efficiency increases slowly with increasing jet $p_{\rm T}$ and has a weak multiplicity dependence. 

The relative systematic uncertainty on track momentum resolution was estimated from the study of the invariant mass distributions of $\Lambda$ and $\mathrm{K^0_s}$ as a function of $\pt$ and amounts to 20\%~\cite{ref:resolutionSys}. 
This track $\pt$ resolution uncertainty was then propagated to the corrected jet spectra with a similar method as used for the tracking efficiency uncertainty evaluation.
The resulting uncertainty from track $\pt$ resolution is about 2\%.

In order to assign the uncertainty arising from the unfolding corrections, several variations are considered. 
By default, the reconstructed jet spectra were unfolded using a detector response matrix and prior spectrum obtained based on events generated by the PYTHIA8 generator with the Monash 2013 tune~\cite{Pythia8Monash}. 
The prior spectrum is used as initial guess of the true spectrum. 
The dependence on MC event generator was quantified by comparing the spectra unfolded using the response matrix and prior from the default generator with those unfolded with response from the EPOS generator with LHC tune~\cite{PhysRevC.92.034906}. The resulting uncertainty is of the order of 3\%. 
Second, the SVD unfolding method~\cite{A44_unfold-svd}, as the default approach used in this paper, was regularised by the choice of parameter $k$ for each cone radius parameter. To estimate the related systematic uncertainty, the regularisation parameter was varied by $\pm 2$ around the optimised value. 
The unfolded results were stable against regularisation parameter variations, and the corresponding systematic uncertainty is negligible.
To validate the unfolding procedure, the SVD unfolded spectra were compared with the results obtained with the Bayesian unfolding method and the remaining differences were taken as an uncertainty. 
In addition to the above variations, the bin boundary migration uncertainty was also evaluated by changing the boundaries of the input spectra and the response matrix during the unfolding process. 
The uncertainties discussed above were then with the RMS calculated for all variations and referred to as the unfolding systematic uncertainty in Table~\ref{tab:inclusivesys} and Table~\ref{tab:multsys}.

The systematic uncertainty due to the background fluctuation estimation was quantified by comparing unfolded spectra obtained with $\delta p_{\rm T}$ distributions using the method of RC without two leading jets (default) and the track embedding method as discussed in Sec.~\ref{subsec:reco}.

The difference between both corrected jet spectra was assigned as background fluctuation uncertainty.
A systematic uncertainty on the background density $\rho_{\rm ch}$ measurement was estimated to be 5\%, resulting in a 2\% uncertainty on the UE-subtracted jet cross section at $\pt = 5\ \mathrm{GeV}/c$, and smaller for higher jet transverse momentum. 
This uncertainty is highest at low $p_{\mathrm{T,jet}}$ in high multiplicity events. 

By default the multiplicity percentiles were determined from the measured distribution of V0M amplitude in data as listed in Table~\ref{tab:V0Mmultiplicity}. 
As shown in Fig.~\ref{fig:V0Mmultiplicity}, PYTHIA8 MC cannot reproduce such multiplicity distribution, which is mainly attributed to a limited description of the UE~\cite{pp13TeV_UnderlyingEvent_ALICE,UErhoCMS7tev,ALICE:2011ac} and inaccurate description of secondary particles and magnetic field during particle transport in the detector simulation. To prevent such multiplicity differences from being propagated as multiplicity dependent results, the unfolding corrections use the data-driven approach with
the background fluctuation response matrices taken from the corresponding event multiplicity class based on data directly, see Fig.~\ref{fig:deltapt}.
This matrix was multiplied with the instrumental matrix obtained from PYTHIA minimum bias events.
To account for the multiplicity estimation uncertainty, a response matrix obtained from pure MC simulation was also used, where the multiplicity intervals and the background fluctuations were both extracted from PYTHIA8 generator.
The systematic uncertainty was assigned based on the comparison of the unfolded spectra obtained from the default analysis and this variation.  
Such uncertainty reaches 
5.7\% for low-$\pt$ jets in the 
highest multiplicity class, and decreases in the lower multiplicity percentile intervals. The uncertainty is independent of the jet $R$ since the multiplicity estimation is at the event level,
and the ratio of jet yields of different $R$ is independent of multiplicity. 

Secondary charged particles are mostly produced by weak decays of strange particles ($\mathrm{K^0_s}$, $\Lambda$, etc.), photon conversions, hadronic interactions in the detector material, and decays of charged pions. Contamination from such secondary charged particles was significantly reduced by a requirement on the maximum distance of closest approach (DCA) of the tracks to the primary vertex. Therefore, the systematic uncertainties due to secondaries were estimated by varying the DCA threshold of track selection, resulting in a jet $\pt$ scale uncertainty of 0.5\%~\cite{Acharya:2018eat,ALICE_chJets7TeV}, which turns into a jet cross section uncertainty of about 3\%.

A systematic uncertainty on the integrated luminosity measurement of 2\%~\cite{ALICE-PUBLIC-2021-005} was assigned to the inclusive jet cross section measurement as a normalisation uncertainty which consequently does not affect the ratios of the measured cross section spectra.

When calculating the systematic uncertainties on the ratios of jet spectra, each uncertainty source was varied simultaneously both for numerator and denominator, and the ratio was calculated using the varied spectra. The resulting difference between the varied spectra and the nominal one is taken as the uncertainty on the ratio for that given uncertainty source. 
This results in a significant 
reduction of the correlated uncertainties from the cancellation between the numerator and the denominator~\cite{Abelev:2013kqa}. The remaining relative difference from each source was added in quadrature. 

The statistical uncertainties on the jet production ratio were also treated carefully between 
numerator and denominator. To avoid the statistical correlations, the total event sample was divided into two parts for the calculation of the numerator and denominator, respectively. The resulting statistical uncertainty on the ratio remains smaller than the systematic uncertainty.

\section{Results}   
\label{sec:results}
\subsection{Inclusive jet cross section measurements}

\begin{figure*}[tbp!]
 \begin{center}   
   \includegraphics[width=0.6\textwidth]{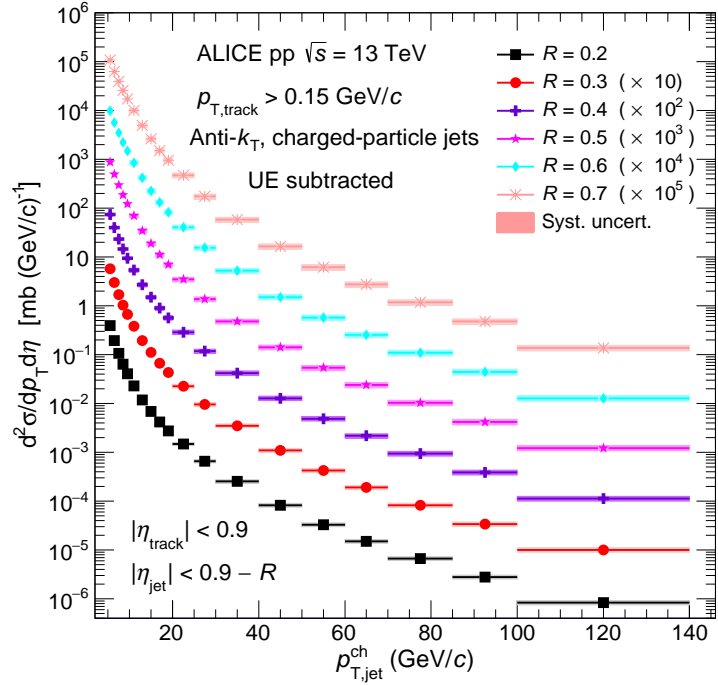}   
 \end{center}
 \caption{Inclusive charged-particle jet cross sections in pp collisions at $\sqrt{s} = 13$ TeV using the anti-$k_{\rm T}$ algorithm for different jet resolution parameters $R$ from $0.2$ to $0.7$, with UE subtraction. Statistical uncertainties are displayed as vertical error bars. The total systematic uncertainties are shown as solid boxes around the data points.}
 \label{fig:CSwUE}
\end{figure*}

The fully-corrected inclusive charged-particle jet cross sections after UE subtraction in pp collisions at $\sqrt{s}=13~\mathrm{TeV}$ are shown in Fig.~\ref{fig:CSwUE} as a function of jet $\pt$ for jet resolution parameters ranging from $R = 0.2$ to $R = 0.7$ and pseudorapidity ranges $|\eta_{\rm jet}|<0.9 - R$. 
The choice of $R$ changes the relative strength of perturbative and non-perturbative (hadronisation and UE) effects on the jet transverse momentum distribution 
~\cite{Dasgupta:2007wa}.
To be consistent with the multiplicity dependent results, all figures presented hereafter are obtained with UE subtraction, while the same measurements without UE subtraction are listed in Appendix~\ref{app:noUE}.  

\begin{figure*}[htbp!]
 \begin{center}
  \includegraphics[width=1.0\textwidth]{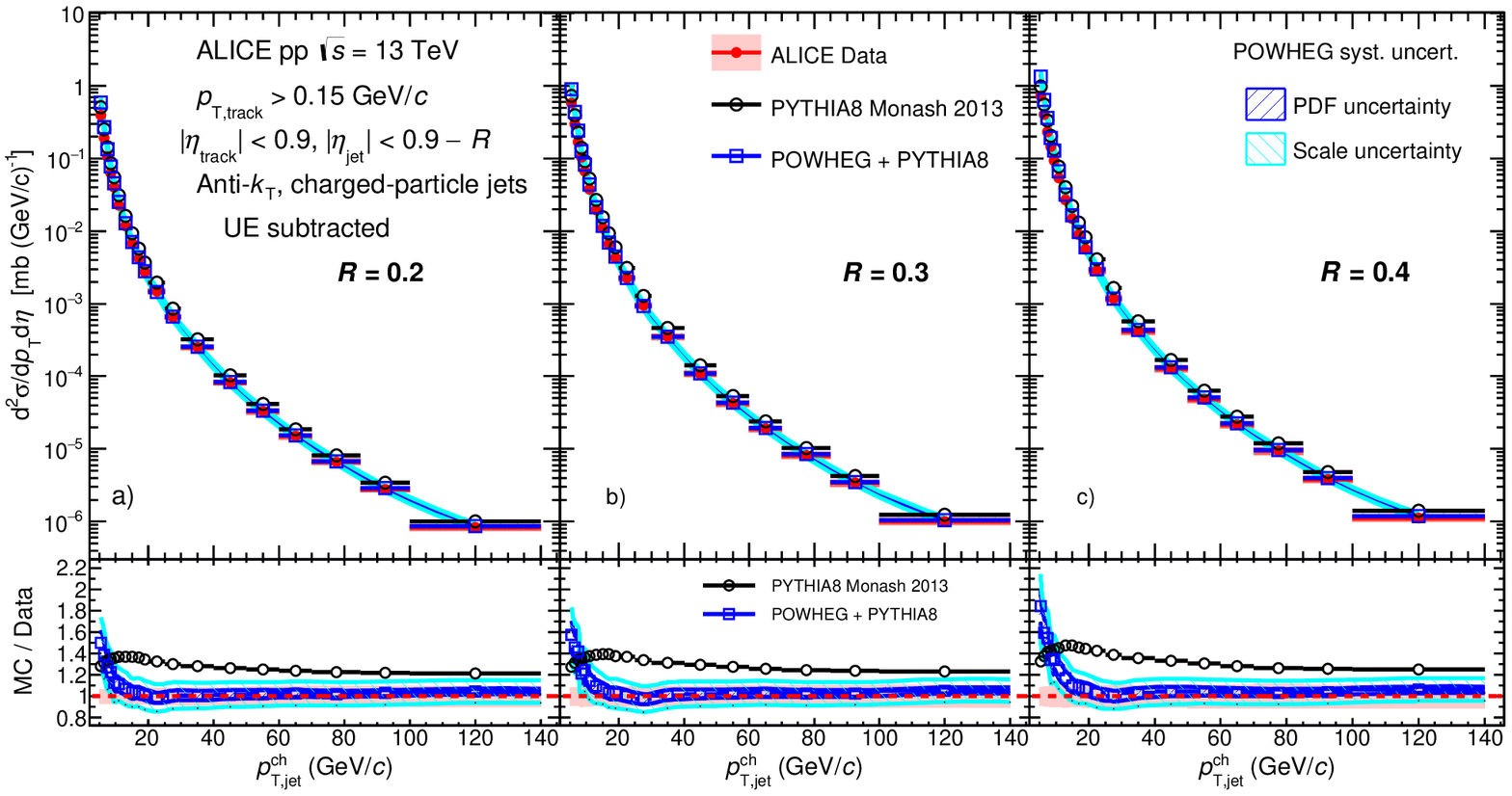} 
  \includegraphics[width=1.0\textwidth]{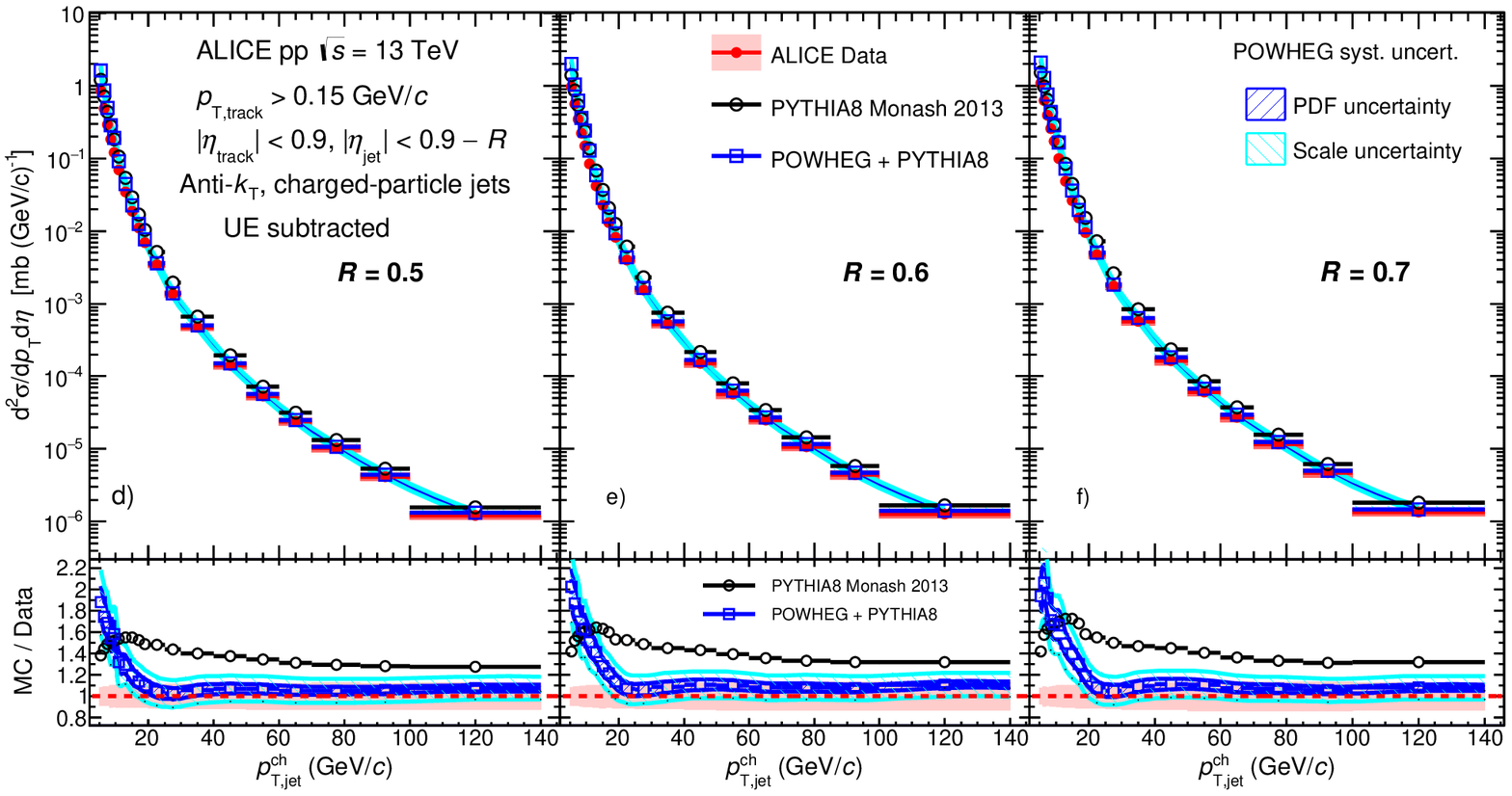}  
 \end{center}
 \caption{Inclusive charged-particle jet cross sections in pp collisions at $\sqrt{s} = 13$ TeV with UE subtraction. Data for different jet resolution parameters $R$ varied from $0.2$ to $0.7$ are compared to LO and NLO MC predictions. The statistical uncertainties are displayed as vertical error bars. The systematic uncertainties on the data are indicated by shaded boxes in the top panels and shaded bands drawn around unity in the bottom panels. The red lines in the ratio correspond to unity.}
 \label{fig:InclusiveJetCS}
\end{figure*}

Figure~\ref{fig:InclusiveJetCS} compares the inclusive charged-particle jet cross sections with predictions from the PYTHIA8 and POWHEG MC event generators after UE subtraction, with the same selections and background subtraction procedure applied. 
The ratios of the MC simulations to ALICE data are shown in the bottom panels of Fig.~\ref{fig:InclusiveJetCS}. The POWHEG MC provides a better description of the data within uncertainties for $p_\mathrm{T,jet}^{\rm ch}\gtrsim 20\ \mathrm{GeV}/c$. Nevertheless, large deviations occur for jet transverse momenta below $20\ \mathrm{GeV}/c$ where POWHEG overestimates the data. Such deviation increases with the jet $R$. A similar enhancement of POWHEG with respect to the data is also observed at 7 TeV~\cite{Acharya:2018eat}, where the implementation of MPI in PYTHIA shows a significant effect on the low $\pt$ jet yield when coupled with POWHEG. 

\begin{figure*}[htbp!]
 \begin{center}   
   \includegraphics[width=0.8\textwidth]{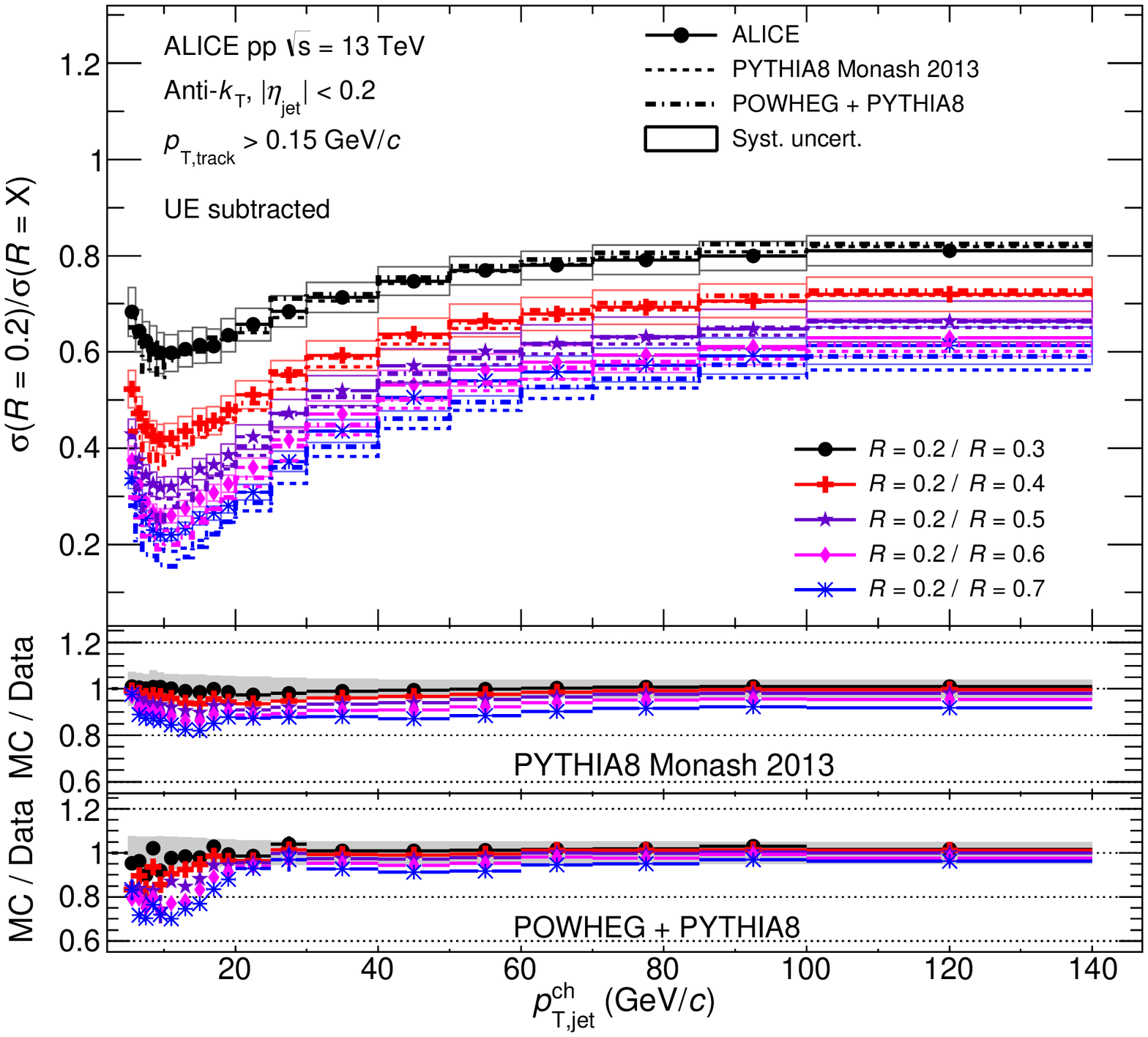}   
 \end{center}
 \caption{Ratio of charged-particle jet cross section for resolution parameter $R = 0.2$ to other radii $R = X $, with $X$ ranging from $0.3$ to $0.7$, after UE subtraction. Data are compared with LO (PYTHIA) and NLO (POWHEG+PYTHIA8) predictions as shown in the bottom panels. The systematic uncertainties of the cross section ratios from data are indicated by solid boxes around data points in the upper panel and shaded bands around unity in the mid and lower panels. No uncertainties are shown for theoretical predictions for better visibility.}
 \label{fig:CSRatio}
\end{figure*}

Figure~\ref{fig:CSRatio} shows the inclusive jet cross section ratios for jets reconstructed with a resolution parameter of $R = 0.2$ to other resolution parameters $R = 0.3$ to $0.7$. 
The observable defined by the ratio of inclusive jet cross sections relates directly to the relative difference between jet $\pt$ distributions when using different resolution parameters and therefore provides insights into the angular dependence of jet fragmentation. 

This observable is also less sensitive to experimental systematic uncertainties since 
the correlated uncertainty on the numerator and denominator spectra are largely cancelled in the ratio. Consequently, the comparisons between data and model predictions 
provide better precision than those for inclusive spectra.
In order to compare the ratios within the same jet pseudorapidity range, the ratios were studied for jet $|\eta_{\rm jet}| < 0.2$, which coincides with the fiducial jet acceptance for the largest resolution parameter studied ($R = 0.7$). 
Statistical correlations between the numerator and denominator of the jet cross section ratios were removed by using exclusive subsets events for their respective assessments.
The measured ratios were compared with PYTHIA and POWHEG calculations in Fig.~\ref{fig:CSRatio}.
Both predictions give a reasonable description of the data for high-$\pt$ jets within 10\%, although they fail to describe the low-$\pt$ region, especially for large resolution parameters, where non-perturbative and UE contributions become large.  
Even though PYTHIA8 overestimates the jet yields (see Fig.~\ref{fig:InclusiveJetCS}), the jet production ratio can still be well described by PYTHIA8 MC.

\begin{figure*}[htbp!]
 \begin{center}
   \includegraphics[width=0.49\textwidth]{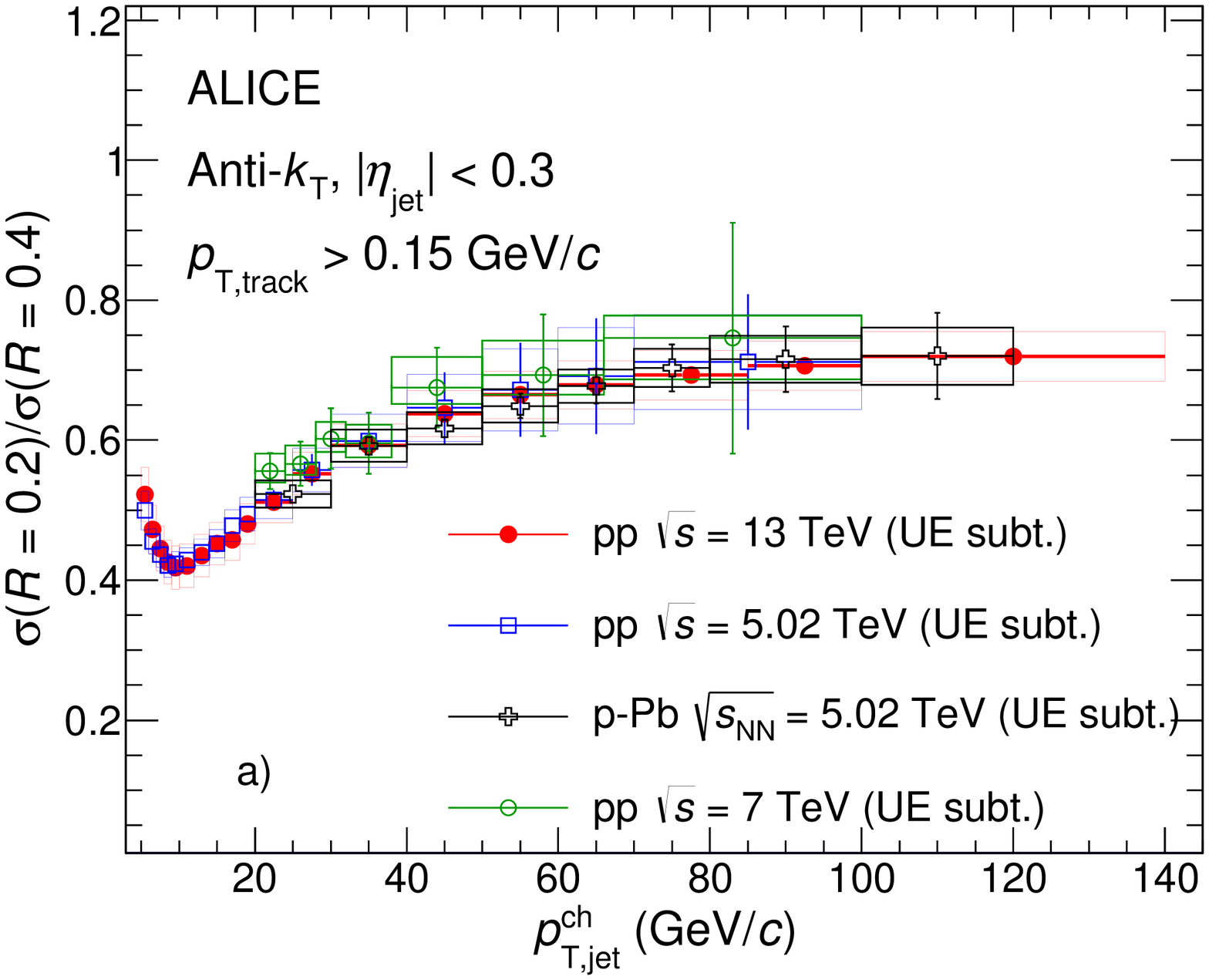} 
   \includegraphics[width=0.49\textwidth]{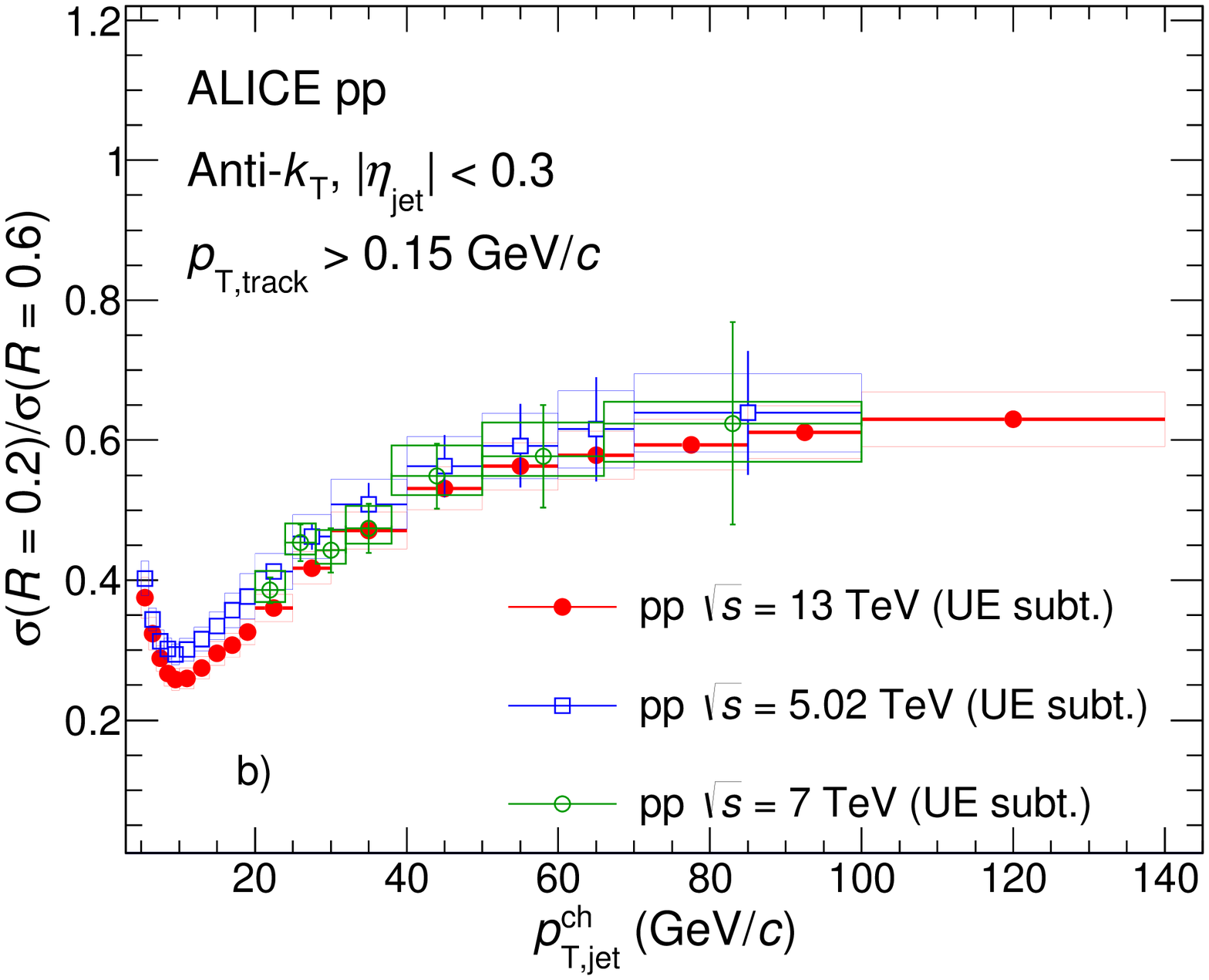}        
 \end{center}
 \caption{Comparison of charged-particle jet cross section ratio with UE subtraction in pp collisions at $\sqrt{s} = 5.02$~\cite{Acharya:2019tku}, $7$~\cite{ALICE_chJets7TeV}, and $13$  $\mathrm{TeV}$ and in p--Pb collisions at $\sqrt{s_\mathrm{NN}} = 5.02 \ \mathrm{TeV}$~\cite{Adam:2015hoa}. Results are a) $\sigma(R=0.2)/\sigma(R=0.4)$ , and b) $\sigma(R=0.2)/\sigma(R=0.6)$.}
 \label{fig:CSRatioComp}
\end{figure*}

Figure~\ref{fig:CSRatioComp} shows the ratio of the charged-particle jet cross section with different $R$ values for 
a) $R=0.2 / R=0.4$ and b) $R=0.2 / R=0.6$ in pp collisions at $\sqrt{s} = 5.02$~\cite{Acharya:2019tku}, $7$~\cite{ALICE_chJets7TeV}, $13$~$\mathrm{TeV}$, and p--Pb collisions at $\sqrt{s_\mathrm{NN}} = 5.02 \ \mathrm{TeV}$~\cite{Adam:2015hoa}.
These results, which are in good agreement within uncertainties, show a similar increase of the jet cross section ratio as a function of jet $\pt$, as expected from the stronger collimation of high-$\pt$ jets. 
No significant energy nor collision species dependence is observed within uncertainties.

\subsection{Multiplicity dependence of jet production}
\begin{figure*}[htbp!]
\begin{center}
\includegraphics[width=1.0\textwidth]{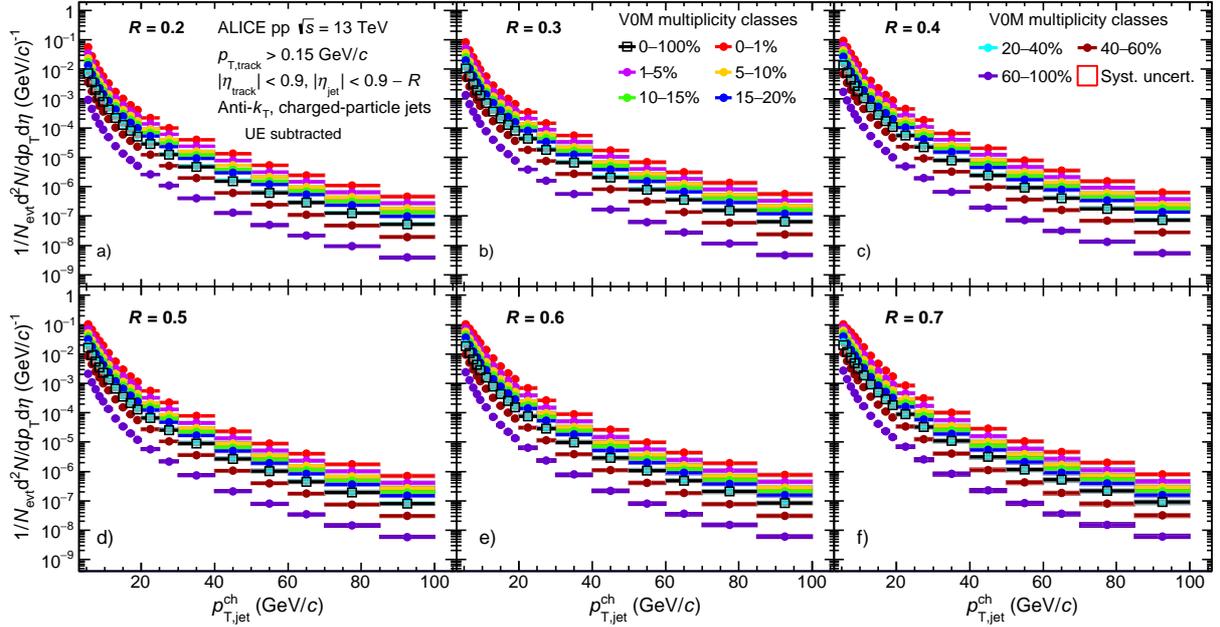} 
\end{center}
\caption{Charged-particle jet yields in different V0M multiplicity percentile intervals for resolution parameters $R$ varied from $0.2$ to $0.7$ in pp collisions at $\sqrt{s} = 13\ \mathrm{TeV}$. Statistical and total systematic uncertainties are shown as vertical error bars and boxes around the data points, respectively.}
\label{fig:MultJetProduction}
\end{figure*}

\begin{figure*}[htbp!]
\begin{center}
\includegraphics[width=1.0\textwidth]{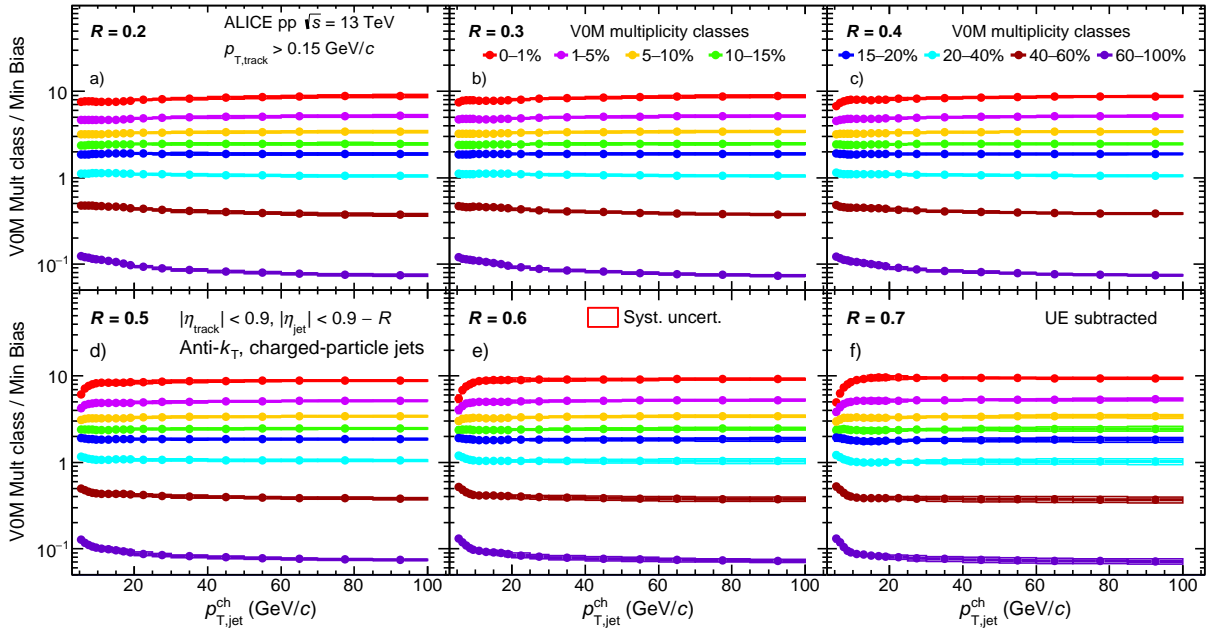} \end{center}
\caption{Ratio of charged-particle jet yield measured in different multiplicity classes with respect to that in MB events as a function of $p_{\rm T}$ for different resolution parameters $R$ from $0.2$ to $0.7$. Statistical and total systematic uncertainties are shown as vertical error bars and boxes around the data points, respectively.
}
\label{fig:MultJetProductionRatio}
\end{figure*}

The jet production yields measured in different V0M multiplicity intervals as a function of jet $p_\mathrm{T}$ for different resolution parameters $R$ varied from $0.2$ to $0.7$ in pp collisions at $\sqrt{s} = 13\ \mathrm{TeV}$ are shown in Fig.~\ref{fig:MultJetProduction}. A higher (lower) jet yield is observed in higher (lower) multiplicity classes. To better investigate this multiplicity dependence, the ratios of jet spectra from multiplicity classes and with MB events 
(Fig.~\ref{fig:CSwUE}) are presented in Fig.~\ref{fig:MultJetProductionRatio}.
The charged-particle jet yield ratio in the highest event multiplicity class (0--1\%) is about 10 times higher than in the MB case, independent of the jet resolution parameter $R$, while in the lowest event class (60--100\%), it amounts to only about 10\% of the MB yield. Furthermore, such ratio has a weak $p_\mathrm{T}$ dependence, except for the very low $p_\mathrm{T}$ region. This indicates that jet production changes with event activity, but the slope of the spectrum stays similar to the one measured in MB events.

\begin{figure*}[htbp!]
 \begin{center}
 \includegraphics[width=1.0\textwidth]{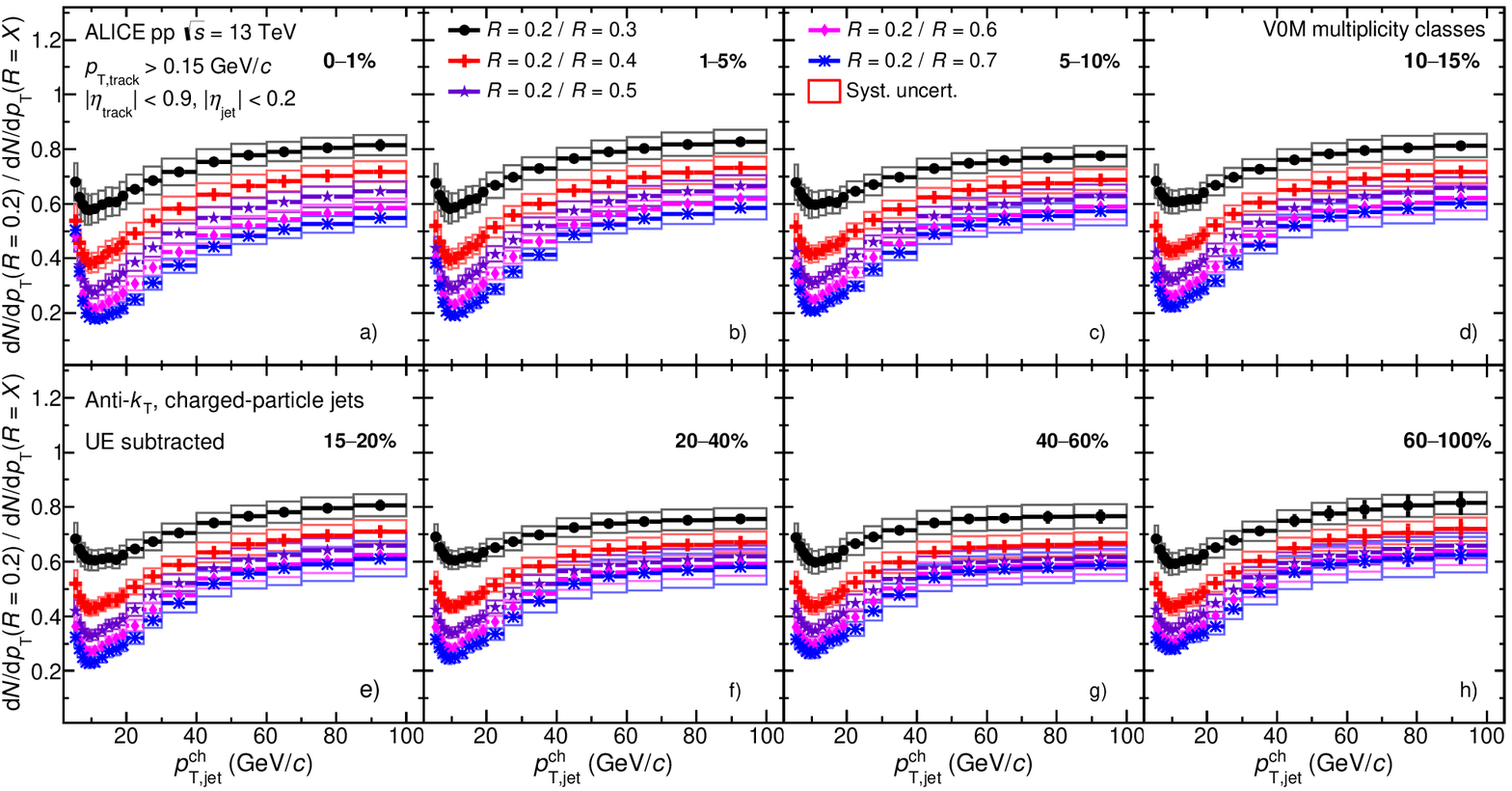} 
 \end{center}
 \caption{Ratios of charged-particle jet spectra with $R = 0.2$ to that with other jet resolution parameters $R$ from $0.3$ to $0.7$, shown in different V0M multiplicity classes. Statistical and systematic uncertainties are shown as vertical error bars and boxes around the data points, respectively.}
 \label{fig:MultJetCSRatio}
\end{figure*}

\begin{figure*}[htbp!]
 \begin{center}
 \includegraphics[width=1.0\textwidth]{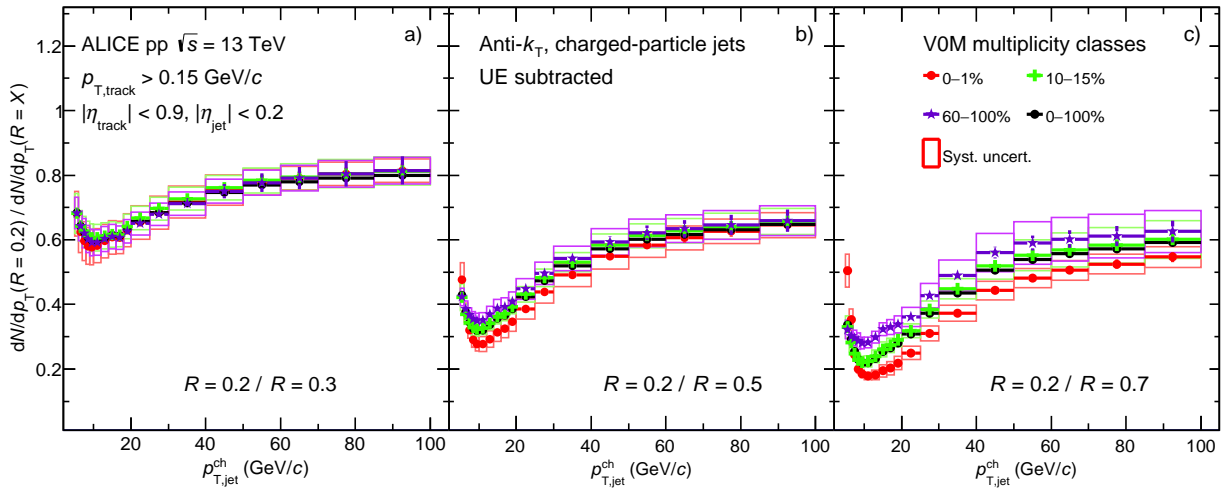}
 \end{center}
 \caption{Comparison of jet spectra ratios of $R = 0.2$ to other radii $R = $ a) $0.3$, b) $0.5$, c) $0.7$ in MB events and in three multiplicity intervals (0--1\%, 10--15\% and 60--100\%). Statistical and systematic uncertainties are shown as vertical error bars and boxes around the data points, respectively. 
 Results for other radii can be found in Appendix Fig.~\ref{fig:MultJetCSRatioCompData02046}.}
 \label{fig:MultJetCSRatioCompData020357}
\end{figure*}

\begin{figure*}[htbp!]
 \begin{center}
 \includegraphics[width=1.0\textwidth]{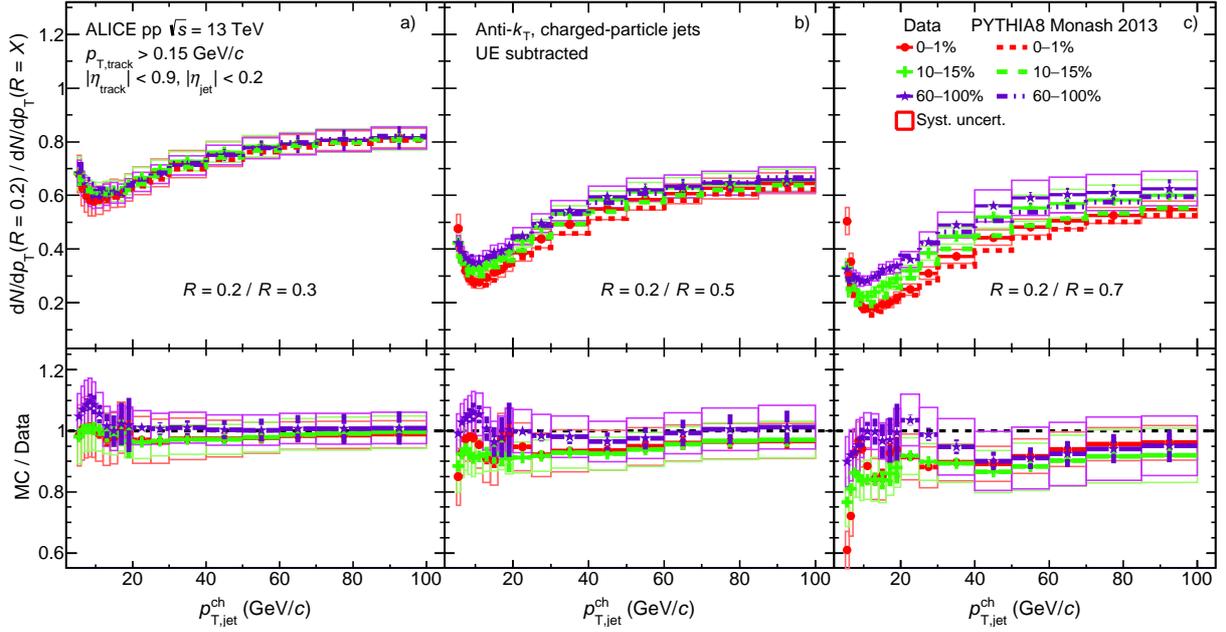}
 \end{center}
 \caption{Comparison of jet spectra ratios of $R = 0.2$ to $R =$ a) $0.3$, b) $0.5$, c) $0.7$ in three multiplicity intervals (0--1\%, 10--15\% and 60--100\%) and compared with PYTHIA8 simulations. Statistical and systematic uncertainties are shown as vertical error bars and boxes around the data points, respectively.
 Results for other radii can be found in Appendix Fig.~\ref{fig:MultJetCSRatioCompMC02046}.}
 \label{fig:MultJetCSRatioCompMC020357}
\end{figure*}

Figure~\ref{fig:MultJetCSRatio} shows the ratios of the $R = 0.2$ jet spectrum to other radii for different multiplicity classes. To better understand the multiplicity dependence of the jet spectra ratios, Fig.~\ref{fig:MultJetCSRatioCompData020357} compares these ratios observed in three selected multiplicity intervals (0--1\%, 10--15\% and 60--100\%) to the ones measured in MB events for a) $R = 0.2/0.3$, b) $0.2/0.5$, and c) $0.2/0.7$. The ratios are consistent with the ones obtained in the MB case (Fig.~\ref{fig:CSRatio}) for small jet radii. At larger jet radii, 
a hint of ordering of the jet production ratios with event multiplicity is observed. It is more pronounced for large radii ($R = 0.2/0.7$) and low $p_{\rm T}$. 
However, with the current systematic uncertainty on data, it is difficult to draw final conclusions on such dependence.
Similar behaviour is observed in MC simulations as shown in Fig.~\ref{fig:MultJetCSRatioCompMC020357}. The MC predictions tend to underestimate the data and this discrepancy increases with jet radius.
However, within the current experimental systematic uncertainties, there is no clear indication of multiplicity dependence for jet yield ratios. 

\begin{figure*}[htbp!]
\begin{center}
\includegraphics[width=0.49\textwidth]{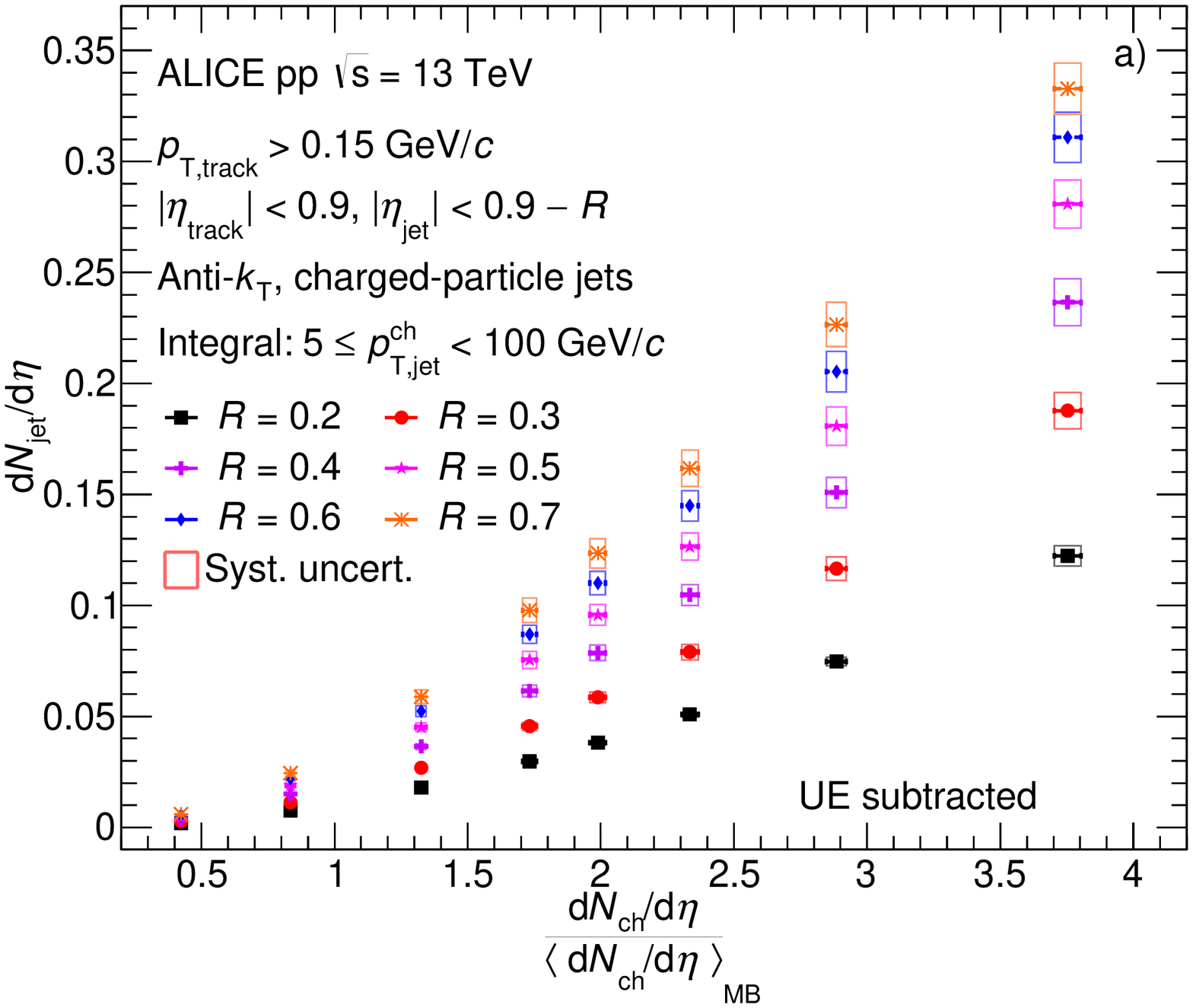}   
\includegraphics[width=0.49\textwidth]{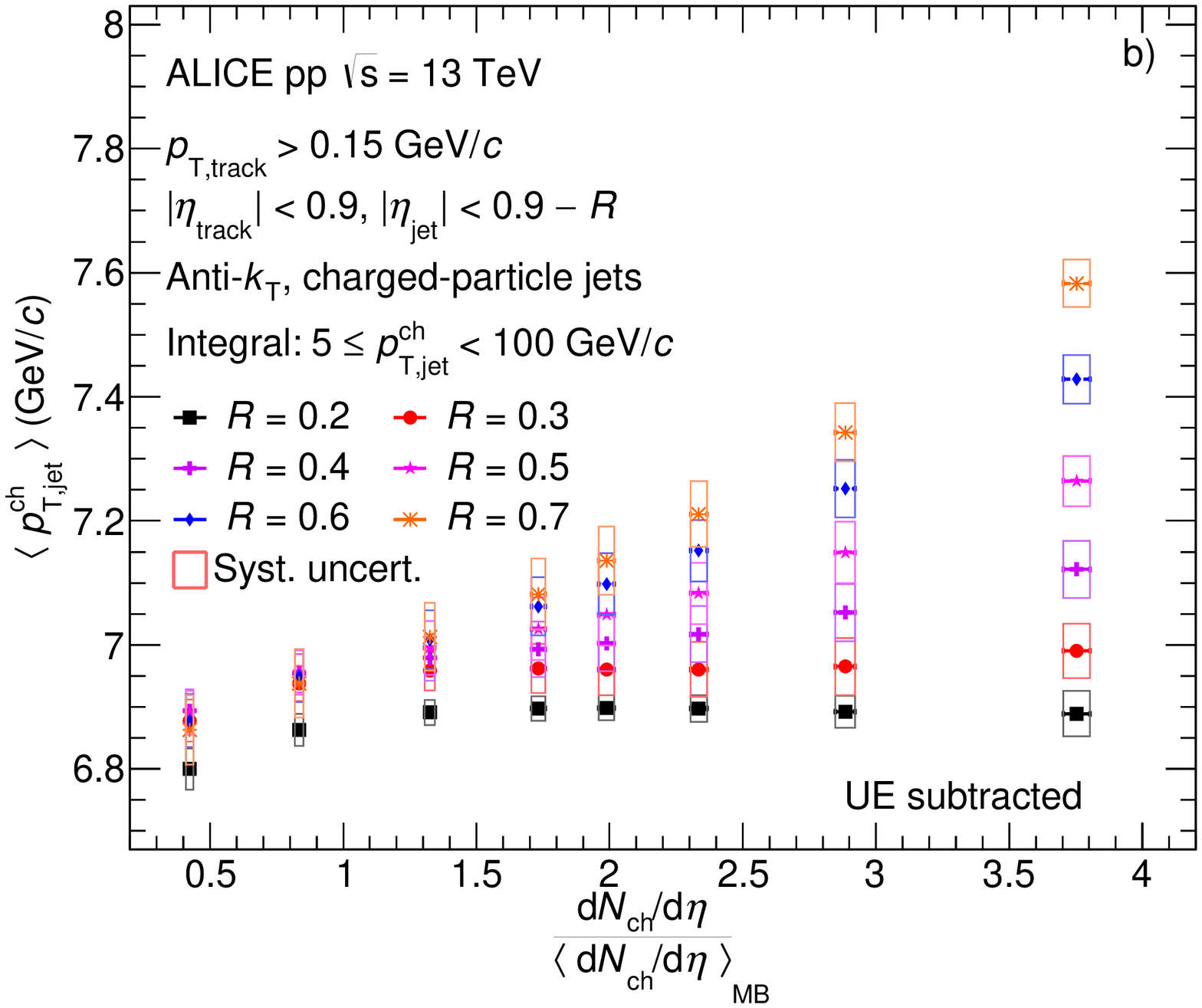}        
\end{center}
\caption{a) Integrated jet yields, and b) $\left \langle p_{\rm T} \right \rangle$ of jets with $5\leq p_{\rm T,jet}^{\rm ch}< 100\ {\rm GeV}/c$ as a function of self-normalised charged-particle multiplicity for different resolution parameters $R$ varied from $0.2$ to $0.7$, with the charged-particle multiplicities provided in Ref.~\cite{Acharya:2020kyh}. Statistical and systematic uncertainties are shown as vertical error bars and boxes around the data points, respectively.}
\label{fig:NjetMean}
\end{figure*}

The $\pt$-integrated ($5\leq p_{\rm T}< 100\ \mathrm{GeV}/c$) jet yields and the average transverse momentum of charged-particle jets as a function of the self-normalised charged-particle multiplicity are shown in Fig.~\ref{fig:NjetMean} for different resolution parameters $R$ from $0.2$ to $0.7$. Both jet yields and the average jet $p_{\rm T}$ increase 
with multiplicity, 
the increase is more evident at larger $R$.
Jets with $R = 0.2$ exhibit very weak dependence of their $\left \langle \pt \right \rangle$ on multiplicity,
indicating that jets reconstructed with small radii are dominated by the leading particle inside in the jet.

\begin{figure*}[htbp!]
\begin{center}
\includegraphics[width=1.0\textwidth]{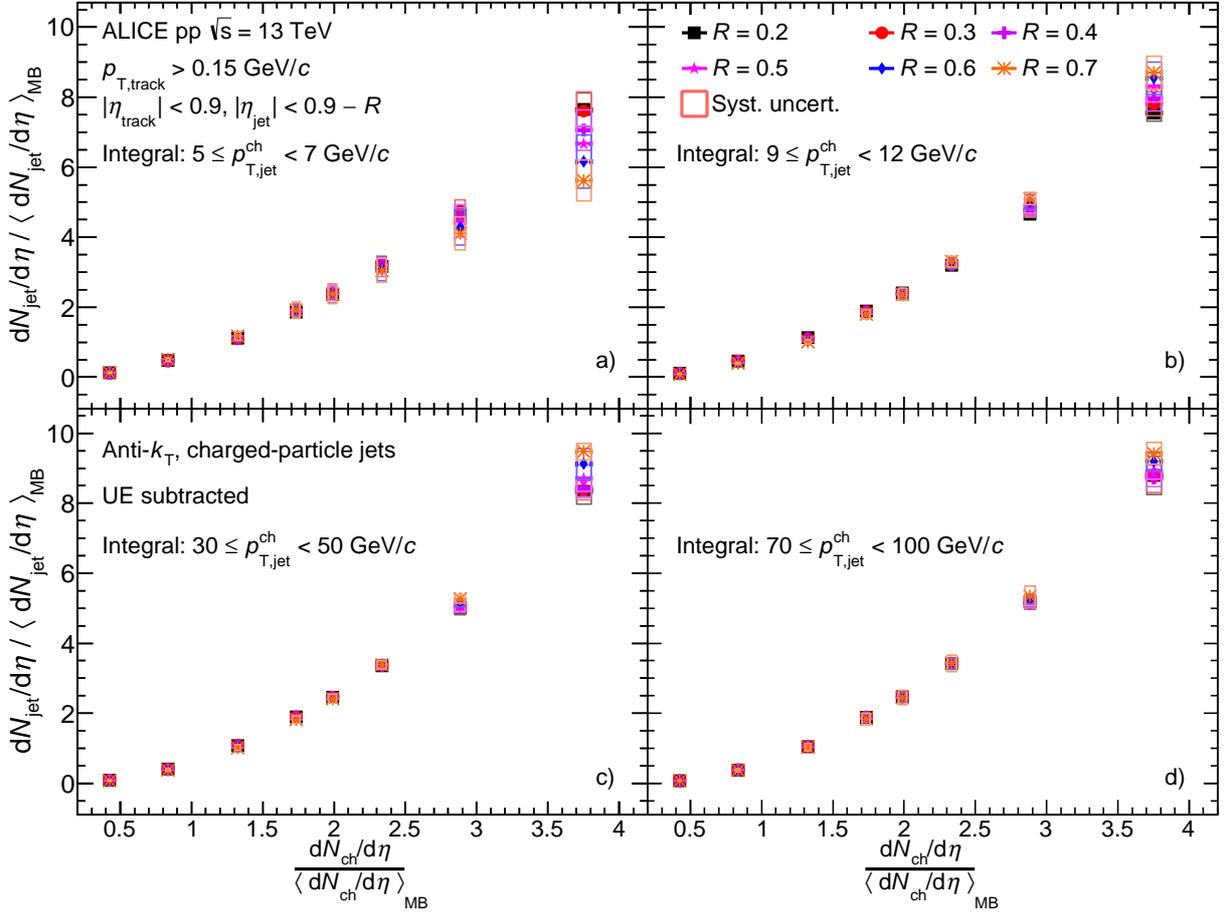}          
\end{center}
\caption{Self-normalised jet yields as a function of the self-normalised charged-particle multiplicity for different resolution parameters $R$ varied from $0.2$ to $0.7$ in different jet $p_{\rm T}$ intervals: a) $5\leq p_{\rm T,jet}^{\rm ch}< 7\ {\rm GeV}/c$, b) $9\leq p_{\rm T,jet}^{\rm ch}< 12\ {\rm GeV}/c$, c) $30\leq p_{\rm T,jet}^{\rm ch}< 50\ {\rm GeV}/c$, and d) $70\leq p_{\rm T,jet}^{\rm ch}< 100\ {\rm GeV}/c$. The charged-particle multiplicities are taken from Ref.~\cite{Acharya:2020kyh}. Statistical and systematic uncertainties are shown as vertical error bars and boxes around the data points, respectively.}
\label{fig:IntegralRatio_R}
\end{figure*}

Fig.~\ref{fig:IntegralRatio_R} presents the jet yield ratios in different multiplicity percentiles with respect to MB events as a function of self-normalised charged-particle multiplicity.
The ratios are shown for four selected jet $\pt$ bins ($5\leq p_{\rm T,jet}^{\rm ch}< 7\ {\rm GeV}/c$, $9\leq p_{\rm T,jet}^{\rm ch}< 12\ {\rm GeV}/c$, $30\leq p_{\rm T,jet}^{\rm ch}< 50\ {\rm GeV}/c$, and $70\leq p_{\rm T,jet}^{\rm ch}< 100\ {\rm GeV}/c$), and for resolution parameters $R = 0.2$\,–\,$0.7$.
The jet yield ratios increase with 
multiplicity for all resolution parameters and $\pt$ intervals. 
No significant dependence of the jet yields with the jet resolution parameter $R$ is seen.

\begin{figure*}[htbp!]
\begin{center}         
\includegraphics[width=1.0\textwidth]{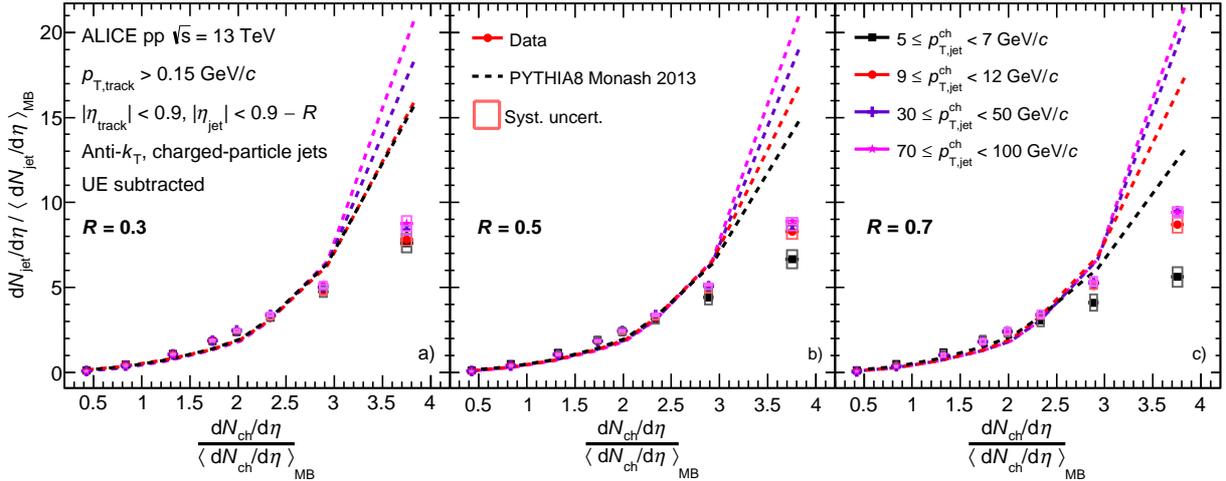}
\end{center}
\caption{Comparison of self-normalised jet yields as a function of the self-normalised charged-particle multiplicity in four selected jet $\pt$ intervals ($5\leq p_{\rm T,jet}^{\rm ch}< 7\ {\rm GeV}/c$, $9\leq p_{\rm T,jet}^{\rm ch}< 12\ {\rm GeV}/c$, $30\leq p_{\rm T,jet}^{\rm ch}< 50\ {\rm GeV}/c$, and $70\leq p_{\rm T,jet}^{\rm ch}< 100\ {\rm GeV}/c$) for a given jet radii: a) $R = 0.3$, b) $R = 0.5$, c) $R = 0.7$ between data and PYTHIA8 predictions, with the charged-particle multiplicities provided in Ref.~\cite{Acharya:2020kyh}. Statistical and systematic uncertainties are shown as vertical error bars and boxes around the data points, respectively. 
 Results for other radii can be found in Appendix Fig.~\ref{fig:IntegralRatio_Pt0246}. }
 \label{fig:IntegralRatio_Pt0357}
\end{figure*}

To explore the $\pt$ dependence of the normalised jet production as a function of self-normalised charged-particle multiplicity, Fig.~\ref{fig:IntegralRatio_Pt0357} shows the self-normalised jet yields as a function of the self-normalised multiplicity 
in four selected jet $\pt$ intervals for resolution parameter $R = 0.3$, $0.5$ and $0.7$. The PYTHIA8 predictions are also compared against data.

The jet production ratios measured at midrapidity increase with multiplicity in a similar way to the results presented in earlier publications for identified particles when using forward multiplicity V0 estimator~\cite{Abelev:2012rz,Acharya:2020pit,Acharya:2020giw}. 
The increase is weaker for the 
lowest jet $\pt$ in the highest multiplicity interval. 
In general, PYTHIA8 simulations predict the overall increasing trend, however, the absolute magnitude is overestimated by the PYTHIA8 MC, especially in the highest multiplicity interval.

\section{Summary} 
\label{sec:conclusion}

The inclusive charged-particle jet production cross sections 
measured
with transverse momentum from $5\ \mathrm{GeV}/c$ to $140\ \mathrm{GeV}/c$ in pp collisions at $\sqrt{s} = 13\ \mathrm{TeV}$ have been reported.
The measurements were performed using the anti-$k_\mathrm{T}$ jet finding algorithm with different resolution parameters $R$ varied from $0.2$ to $0.7$ and the pseudorapidity range $|\eta_{\rm jet}|<0.9 - R$. 
The inclusive charged-particle jet cross sections were compared to LO PYTHIA and NLO POWHEG pQCD calculations. As expected, a better agreement between data and MC is observed for the NLO predictions, although the NLO prediction overestimates the jet yield below $20\ \mathrm{GeV}/c$. The cross section ratios for different resolution parameters were also studied. 
These ratios increase with jet $p_\mathrm{T}$ and saturate at the high end of the jet $p_\mathrm{T}$ range, indicating a stronger collimation for high-momentum jets.

The multiplicity dependence of jet production using different resolution parameters has also been studied. A higher (lower) jet yield is observed in higher (lower) multiplicity classes. Jet production in different multiplicity intervals compared to MB has a weak $p_\mathrm{T}$ and jet resolution parameter dependence. Furthermore, the self-normalised jet production yields and average jet $p_\mathrm{T}$ as a function of the self-normalised charged-particle multiplicity have been measured. The integrated jet yields and $\left \langle \pt \right \rangle$ in the integrated $\pt$ interval between 5 and 100 ${\rm GeV}/c$ increase with the self-normalised charged-particle multiplicity. No strong dependence of jet $p_\mathrm{T}$ and the resolution parameter $R$ are observed except at low transverse momentum in the highest multiplicity percentile interval. A similar multiplicity dependence has also been reported for prompt D mesons in p–Pb collisions at $\sqrt{s_{\rm NN}} = 5.02\ \mathrm{TeV}$ and non-prompt J/$\psi$ (from B hadron decays) production in pp collisions at $\sqrt{s} = 7\ \mathrm{TeV}$ when using a forward multiplicity estimator. Current MC event generators can only predict the rising trend but cannot describe the absolute yields, especially in the highest multiplicity class.

The measurements presented in this paper provide further insight into the interplay between soft particle production and hard processes. Detailed comparisons of models with data will help to elucidate the relationship between jet production mechanisms and high-multiplicity events in small systems, particularly at LHC energies.


\newenvironment{acknowledgement}{\relax}{\relax}
\begin{acknowledgement}
\section*{Acknowledgements}

The ALICE Collaboration would like to thank all its engineers and technicians for their invaluable contributions to the construction of the experiment and the CERN accelerator teams for the outstanding performance of the LHC complex.
The ALICE Collaboration gratefully acknowledges the resources and support provided by all Grid centres and the Worldwide LHC Computing Grid (WLCG) collaboration.
The ALICE Collaboration acknowledges the following funding agencies for their support in building and running the ALICE detector:
A. I. Alikhanyan National Science Laboratory (Yerevan Physics Institute) Foundation (ANSL), State Committee of Science and World Federation of Scientists (WFS), Armenia;
Austrian Academy of Sciences, Austrian Science Fund (FWF): [M 2467-N36] and Nationalstiftung f\"{u}r Forschung, Technologie und Entwicklung, Austria;
Ministry of Communications and High Technologies, National Nuclear Research Center, Azerbaijan;
Conselho Nacional de Desenvolvimento Cient\'{\i}fico e Tecnol\'{o}gico (CNPq), Financiadora de Estudos e Projetos (Finep), Funda\c{c}\~{a}o de Amparo \`{a} Pesquisa do Estado de S\~{a}o Paulo (FAPESP) and Universidade Federal do Rio Grande do Sul (UFRGS), Brazil;
Ministry of Education of China (MOEC) , Ministry of Science \& Technology of China (MSTC) and National Natural Science Foundation of China (NSFC), China;
Ministry of Science and Education and Croatian Science Foundation, Croatia;
Centro de Aplicaciones Tecnol\'{o}gicas y Desarrollo Nuclear (CEADEN), Cubaenerg\'{\i}a, Cuba;
Ministry of Education, Youth and Sports of the Czech Republic, Czech Republic;
The Danish Council for Independent Research | Natural Sciences, the VILLUM FONDEN and Danish National Research Foundation (DNRF), Denmark;
Helsinki Institute of Physics (HIP), Finland;
Commissariat \`{a} l'Energie Atomique (CEA) and Institut National de Physique Nucl\'{e}aire et de Physique des Particules (IN2P3) and Centre National de la Recherche Scientifique (CNRS), France;
Bundesministerium f\"{u}r Bildung und Forschung (BMBF) and GSI Helmholtzzentrum f\"{u}r Schwerionenforschung GmbH, Germany;
General Secretariat for Research and Technology, Ministry of Education, Research and Religions, Greece;
National Research, Development and Innovation Office, Hungary;
Department of Atomic Energy Government of India (DAE), Department of Science and Technology, Government of India (DST), University Grants Commission, Government of India (UGC) and Council of Scientific and Industrial Research (CSIR), India;
Indonesian Institute of Science, Indonesia;
Istituto Nazionale di Fisica Nucleare (INFN), Italy;
Japanese Ministry of Education, Culture, Sports, Science and Technology (MEXT) and Japan Society for the Promotion of Science (JSPS) KAKENHI, Japan;
Consejo Nacional de Ciencia (CONACYT) y Tecnolog\'{i}a, through Fondo de Cooperaci\'{o}n Internacional en Ciencia y Tecnolog\'{i}a (FONCICYT) and Direcci\'{o}n General de Asuntos del Personal Academico (DGAPA), Mexico;
Nederlandse Organisatie voor Wetenschappelijk Onderzoek (NWO), Netherlands;
The Research Council of Norway, Norway;
Commission on Science and Technology for Sustainable Development in the South (COMSATS), Pakistan;
Pontificia Universidad Cat\'{o}lica del Per\'{u}, Peru;
Ministry of Education and Science, National Science Centre and WUT ID-UB, Poland;
Korea Institute of Science and Technology Information and National Research Foundation of Korea (NRF), Republic of Korea;
Ministry of Education and Scientific Research, Institute of Atomic Physics, Ministry of Research and Innovation and Institute of Atomic Physics and University Politehnica of Bucharest, Romania;
Joint Institute for Nuclear Research (JINR), Ministry of Education and Science of the Russian Federation, National Research Centre Kurchatov Institute, Russian Science Foundation and Russian Foundation for Basic Research, Russia;
Ministry of Education, Science, Research and Sport of the Slovak Republic, Slovakia;
National Research Foundation of South Africa, South Africa;
Swedish Research Council (VR) and Knut \& Alice Wallenberg Foundation (KAW), Sweden;
European Organization for Nuclear Research, Switzerland;
Suranaree University of Technology (SUT), National Science and Technology Development Agency (NSDTA), Suranaree University of Technology (SUT), Thailand Science Research and Innovation (TSRI) and National Science, Research and Innovation Fund (NSRF), Thailand;
Turkish Energy, Nuclear and Mineral Research Agency (TENMAK), Turkey;
National Academy of  Sciences of Ukraine, Ukraine;
Science and Technology Facilities Council (STFC), United Kingdom;
National Science Foundation of the United States of America (NSF) and United States Department of Energy, Office of Nuclear Physics (DOE NP), United States of America.    
\end{acknowledgement}

\bibliographystyle{utphys}   
\bibliography{biblio}

\newpage
\appendix
\section{Appendix}
\label{app:appendix}
\subsection{Charged-particle jet cross section and ratios without UE subtraction}
\label{app:noUE}
\begin{figure*}[htbp]
 \begin{center}   
   \includegraphics[width=0.8\textwidth]{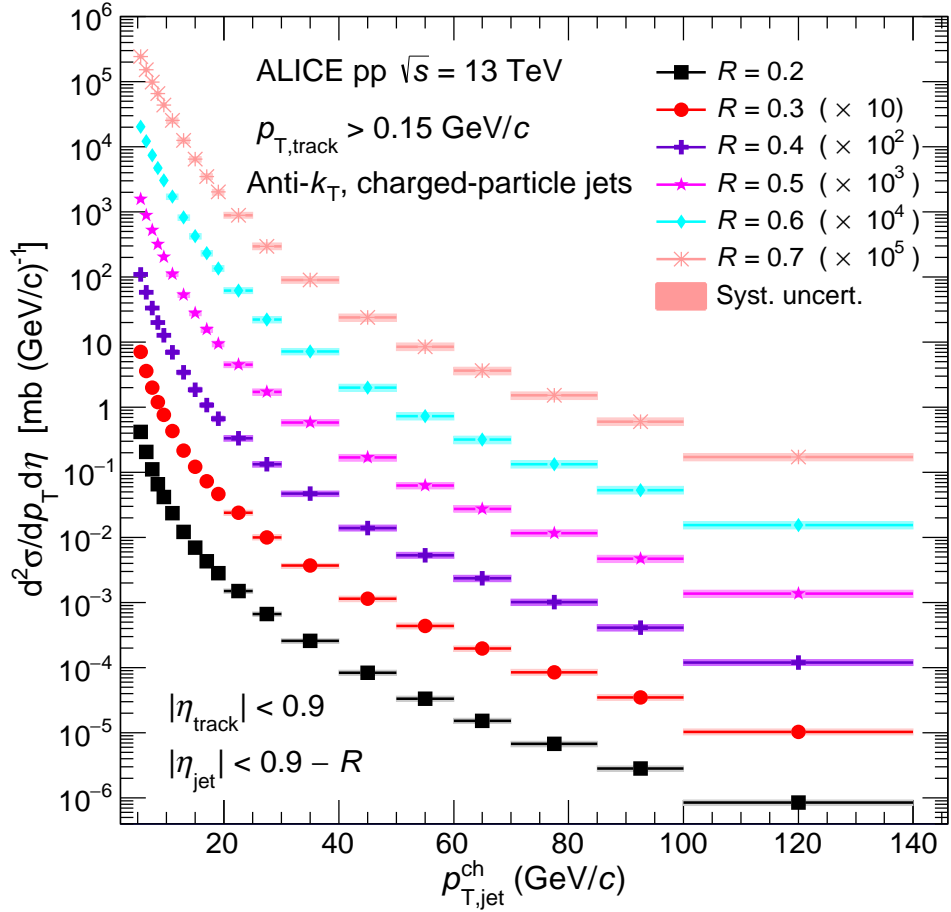} 
 \end{center}
 \caption{Inclusive charged-particle jet cross sections in pp collisions at $\sqrt{s} =$ 13 TeV using the anti-$k_{\rm T}$ algorithm for different resolution parameters $R$ varied from $0.2$ to $0.7$, without UE subtraction. Statistical uncertainties are displayed as vertical error bars. The total systematic uncertainties are shown as solid boxes around the data points.}
 \label{fig:CSwithoutUE}
\end{figure*}

\begin{figure*}[htbp]
 \begin{center}
  \includegraphics[width=1.0\textwidth]{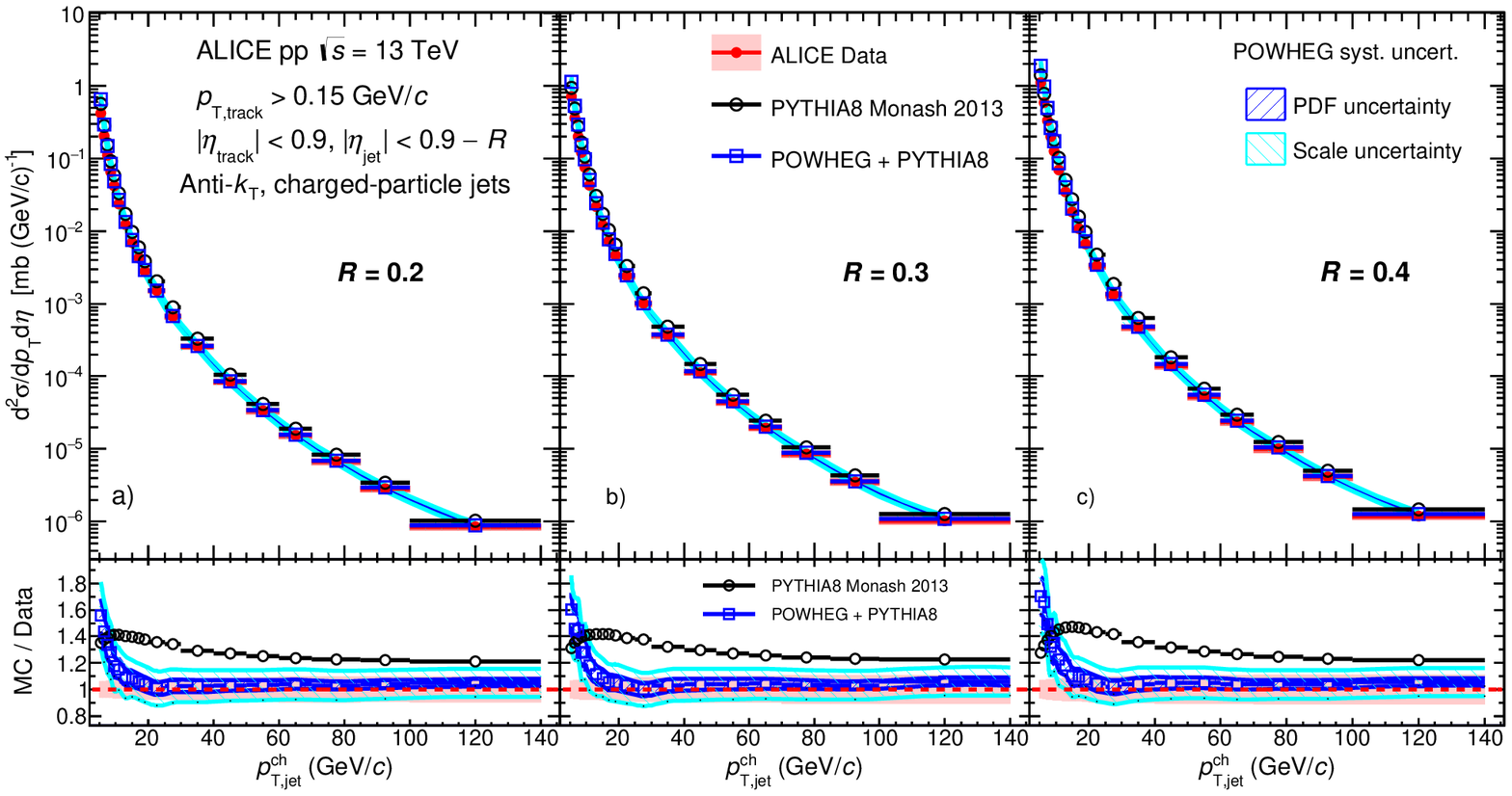} 
  \includegraphics[width=1.0\textwidth]{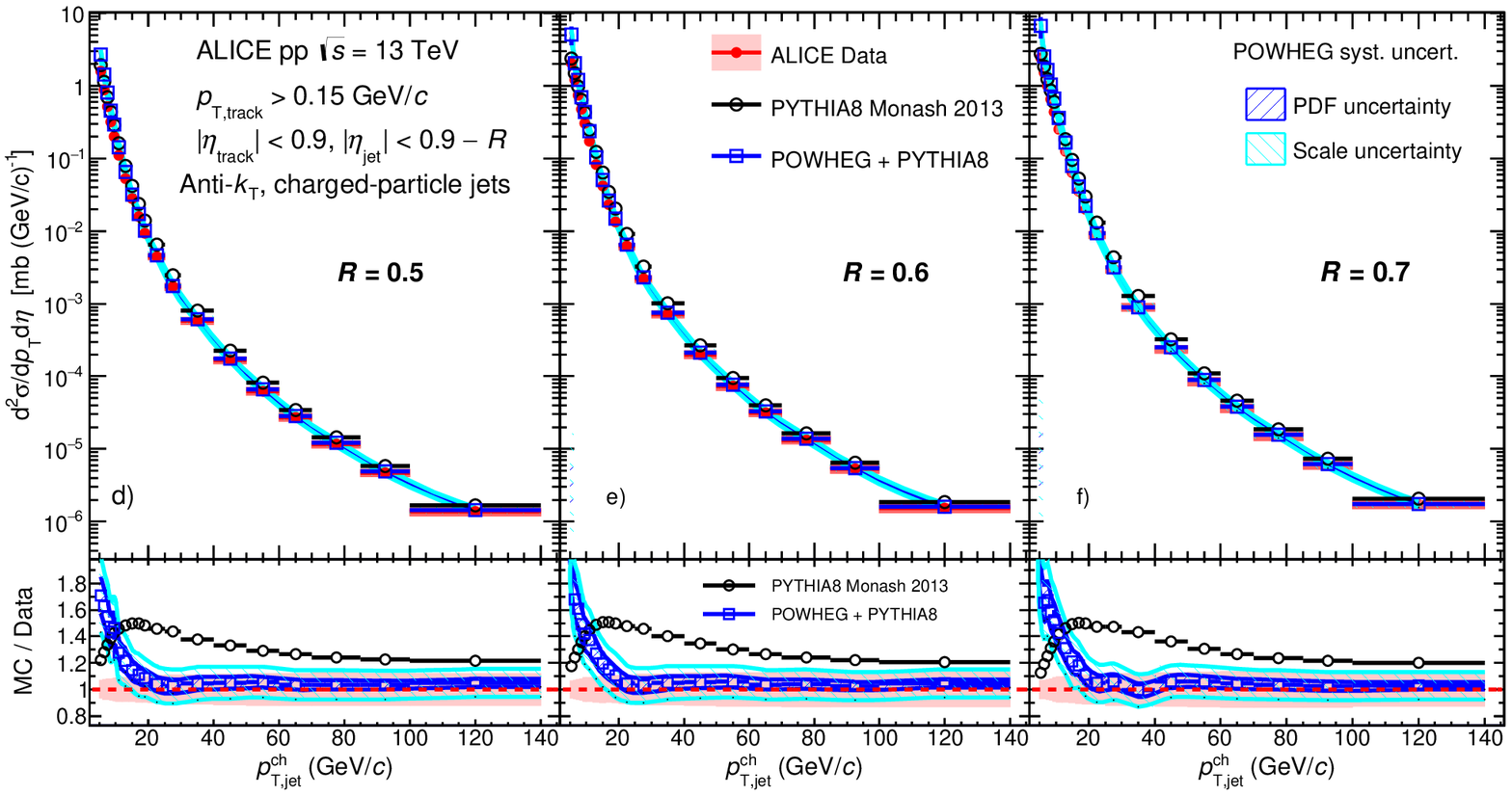}  
 \end{center}
 \caption{Inclusive charged-particle jet cross sections in pp collisions at $\sqrt{s} = 13$ TeV without UE subtraction and compared to LO and NLO MC predictions with different resolution parameters $R$ varied from $0.2$ to $0.7$. The statistical uncertainties are displayed as vertical error bars. The systematic uncertainties on the data are indicated by shaded boxes in the top panels and shaded bands drawn around unity in the bottom panels. The red dashed lines in the ratio correspond to unity.}
 \label{fig:InclusiveJetCSNoUE}
\end{figure*}

\begin{figure*}[htbp]
 \begin{center}   
   \includegraphics[width=0.8\textwidth]{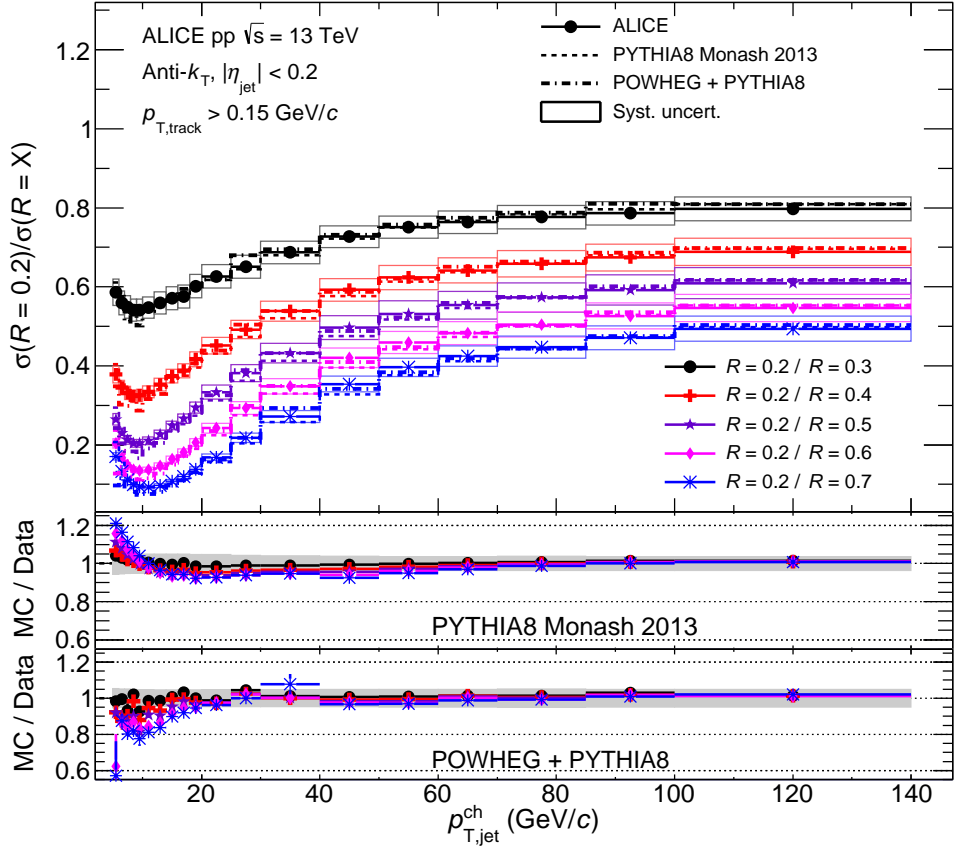} 
 \end{center}
 \caption{Ratio of charged-particle jet cross section for resolution parameter $R = 0.2$ to other radii $R = X $, with $X$ ranging from $0.3$ to $0.7$, without UE subtraction, and the comparison of calculations from LO (PYTHIA) and NLO event generators (POWHEG+PYTHIA8). The systematic uncertainties of the cross section ratios from data are indicated by solid boxes around data points in the upper panels, and shaded bands around unity in the lower panels. No uncertainties are shown for theoretical predictions for better visibility.}
 \label{fig:CSRatioNoUE}
\end{figure*}

\begin{figure*}[htbp]
 \begin{center}
 \includegraphics[width=0.8\textwidth]{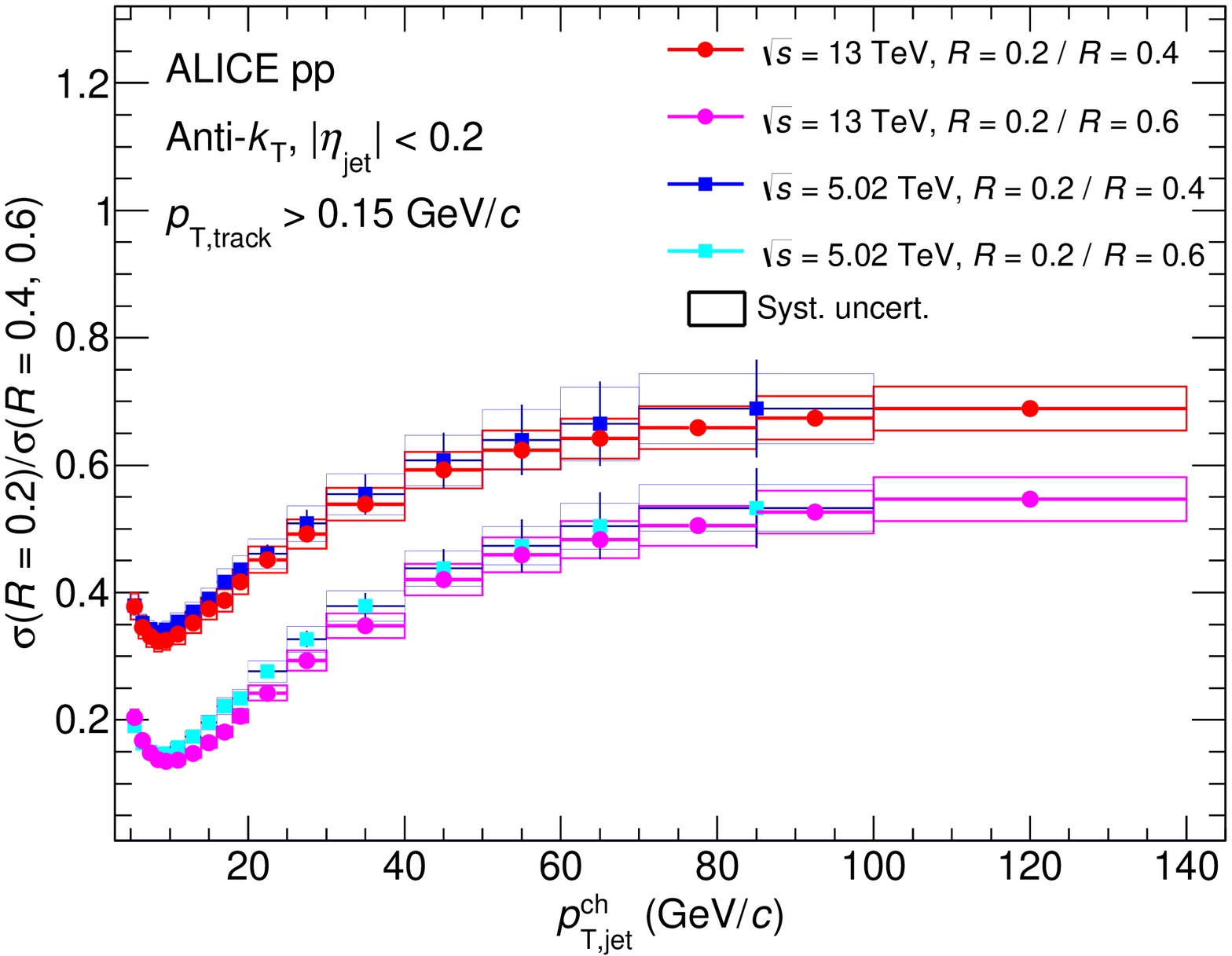}     
 \end{center}
 \caption{Comparison of charged-particle jet cross section ratios for $\sigma(R=0.2)/\sigma(R=0.4)$ and $\sigma(R=0.2)/\sigma(R=0.6)$ without UE subtraction in pp collisions at $\sqrt{s} = 13$ and $5.02$ $\mathrm{TeV}$~\cite{Acharya:2019tku}.}
 \label{fig:CSRatioCompNoUE}
\end{figure*}

The fully corrected inclusive charged-particle jet cross sections and cross section ratios without UE corrections in pp collisions at $\sqrt{s} = 13\ \mathrm{TeV}$ are presented in this section. Figure~\ref{fig:CSwithoutUE} shows the jet cross section for different resolution parameters $R$ varied from $0.2$ to $0.7$ without UE subtraction. The comparisons with LO and NLO theoretical calculations are shown in Fig.~\ref{fig:InclusiveJetCSNoUE}. Figure~\ref{fig:CSRatioNoUE} and~\ref{fig:CSRatioCompNoUE} show the jet cross section ratios without UE subtraction, in addition to comparison with theoretical calculations between different collision energies, respectively. 

\subsection{Multiplicity dependence of jet production}
\begin{figure*}[htbp]
 \begin{center}
  \includegraphics[width=1.0\textwidth]{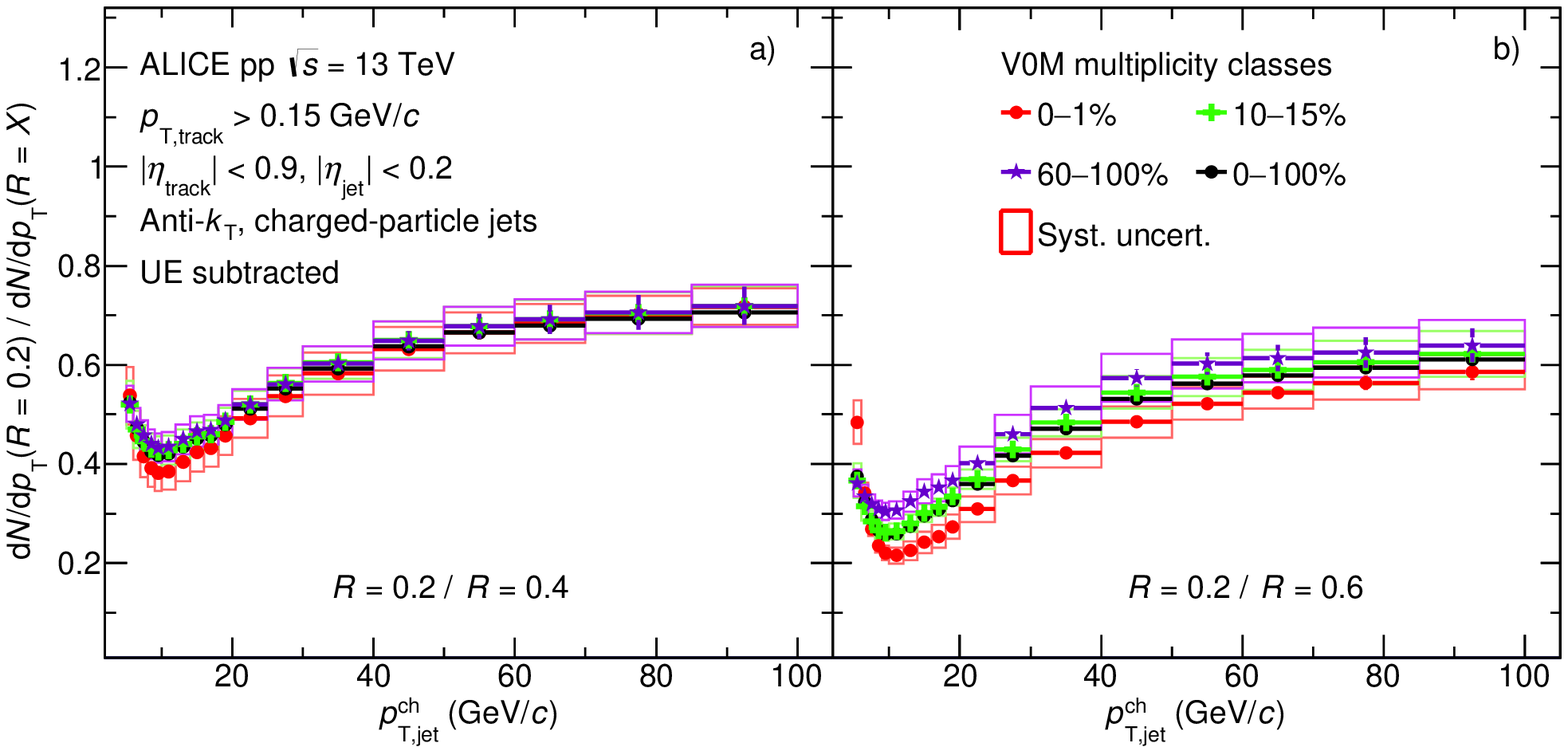}
 \end{center}
 \caption{Comparison of jet production ratios of $R = 0.2$ to a) $R = 0.4$, b) $R = 0.6$ in three multiplicity intervals (0--1\%, 10--15\% and 60--100\%) and compared with PYTHIA simulations. Statistical and systematic uncertainties are shown as vertical error bars and boxes around the data points, respectively.}
 \label{fig:MultJetCSRatioCompData02046}
\end{figure*}

\begin{figure*}[htbp]
 \begin{center}
 \includegraphics[width=1.0\textwidth]{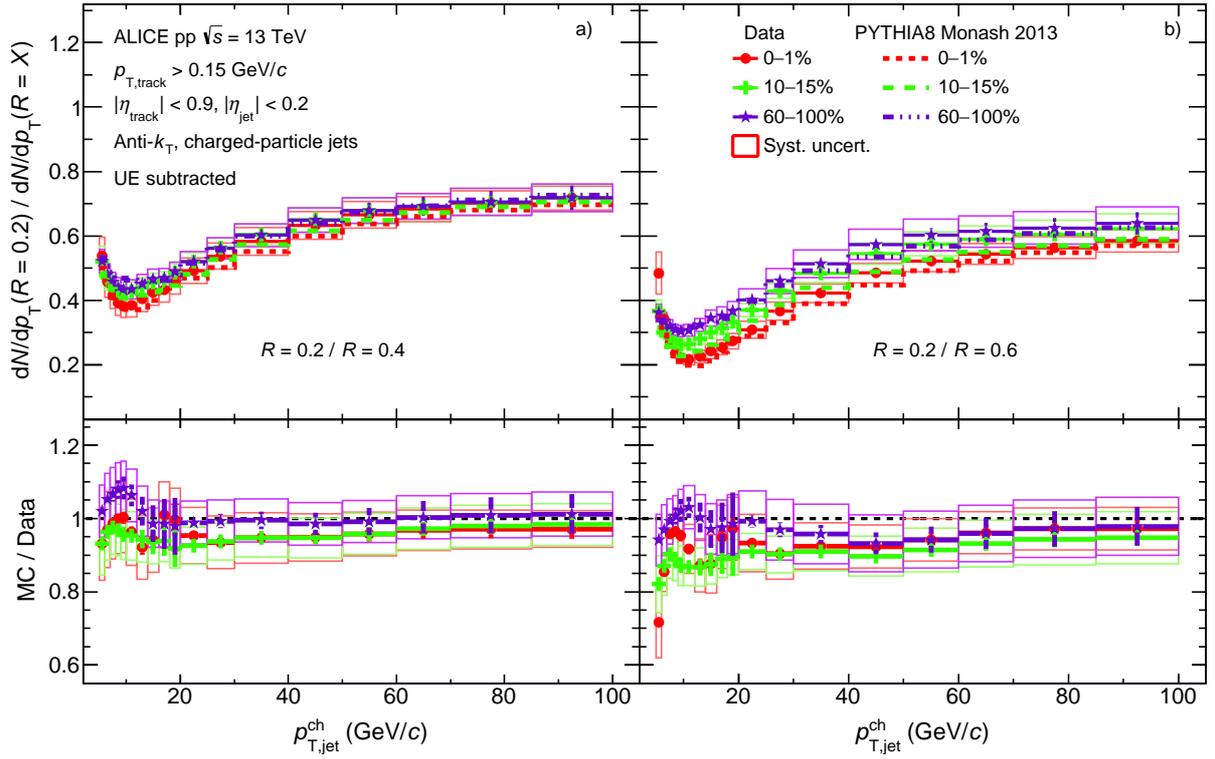}
 \end{center}
 \caption{Comparison of jet production ratios of $R = 0.2$ to a) $R = 0.4$, b) $R = 0.6$ in three different multiplicity classes and compared with PYTHIA MC simulations. Statistical and the total systematic uncertainties are shown as vertical error bars and boxes around the data points, respectively.}
 \label{fig:MultJetCSRatioCompMC02046}
\end{figure*}

\begin{figure*}[htbp]
 \begin{center}         
 \includegraphics[width=1.0\textwidth]{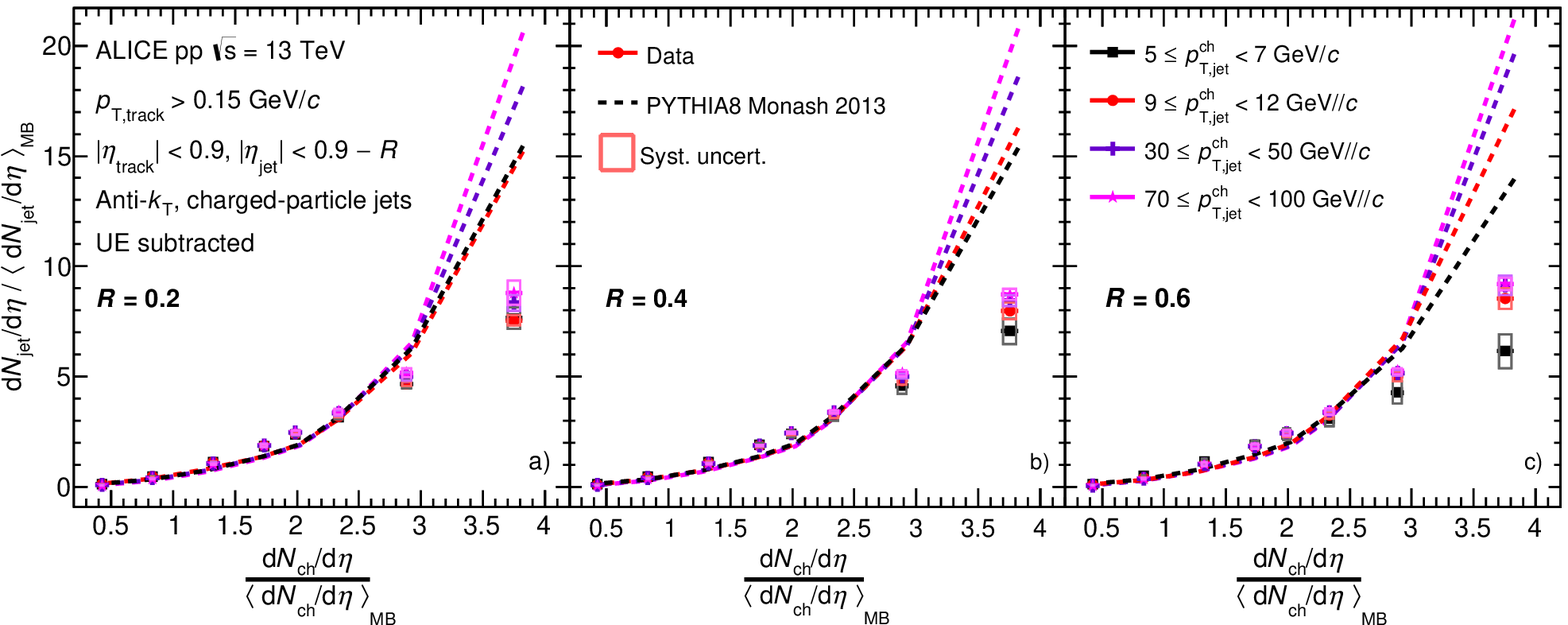} 
 \end{center}
 \caption{Comparison of self-normalised jet yields as a function of the self-normalised charged-particle multiplicity in four selected jet $\pt$ intervals ($5\leq p_{\rm T,jet}^{\rm ch}< 7\ {\rm GeV}/c$, $9\leq p_{\rm T,jet}^{\rm ch}< 12\ {\rm GeV}/c$, $30\leq p_{\rm T,jet}^{\rm ch}< 50\ {\rm GeV}/c$, and $70\leq p_{\rm T,jet}^{\rm ch}< 100\ {\rm GeV}/c$) for a given jet radii: a) $R = 0.2$, b) $R = 0.4$, c) $R = 0.6$ between data and PYTHIA8 predictions. The charged-particle multiplicities are taken from Ref.~\cite{Acharya:2020kyh}. Statistical and systematic uncertainties are shown as vertical error bars and boxes around the data points, respectively.}
 \label{fig:IntegralRatio_Pt0246}
\end{figure*}

Figure~\ref{fig:MultJetCSRatioCompData02046} selects three multiplicity intervals (0--1\%, 10--15\% and 60--100\%) and compares the jet production ratio for a) $R = 0.2/0.4$ and b) $0.2/0.6$.
Then Fig.~\ref{fig:MultJetCSRatioCompMC02046} compares the jet production ratio in three multiplicity classes for a) $R = 0.2/0.4$ and b) $0.2/0.6$ between data and PYTHIA MC, with the ratio between MC and data shown in the bottom panels. 
Figure~\ref{fig:IntegralRatio_Pt0246} shows the jet production ratio as a function of the self-normalised charged-particle multiplicity in four selected jet transverse momentum intervals for jet resolution parameters a) $R = 0.2$, b) $R = 0.4$, and c) $R = 0.6$.

\section{The ALICE Collaboration}
\label{app:collab}
\small
\begin{flushleft} 

S.~Acharya$^{\rm 142}$, 
D.~Adamov\'{a}$^{\rm 96}$, 
A.~Adler$^{\rm 74}$, 
J.~Adolfsson$^{\rm 81}$, 
G.~Aglieri Rinella$^{\rm 34}$, 
M.~Agnello$^{\rm 30}$, 
N.~Agrawal$^{\rm 54}$, 
Z.~Ahammed$^{\rm 142}$, 
S.~Ahmad$^{\rm 16}$, 
S.U.~Ahn$^{\rm 76}$
% ...

\section*{Affiliation Notes}

$^{\rm I}$ Deceased\\
$^{\rm II}$ Also at: Italian National Agency for New Technologies, Energy and Sustainable Economic Development (ENEA), Bologna, Italy\\
$^{\rm III}$ Also at: Dipartimento DET del Politecnico di Torino, Turin, Italy\\
$^{\rm IV}$ Also at: M.V. Lomonosov Moscow State University, D.V. Skobeltsyn Institute of Nuclear, Physics, Moscow, Russia\\
$^{\rm V}$ Also at: Department of Applied Physics, Aligarh Muslim University, Aligarh, India\\
$^{\rm VI}$ Also at: Institute of Theoretical Physics, University of Wroclaw, Poland\\
$^{\rm VII}$ Also at: University of Kansas, Lawrence, Kansas, United States\\

\section*{Collaboration Institutes}

$^{1}$ A.I. Alikhanyan National Science Laboratory (Yerevan Physics Institute) Foundation, Yerevan, Armenia\\
$^{2}$ AGH University of Science and Technology, Cracow, Poland\\
$^{3}$ Bogolyubov Institute for Theoretical Physics, National Academy of Sciences of Ukraine, Kiev, Ukraine\\
$^{4}$ Bose Institute, Department of Physics  and Centre for Astroparticle Physics and Space Science (CAPSS), Kolkata, India\\
$^{5}$ Budker Institute for Nuclear Physics, Novosibirsk, Russia\\
$^{6}$ California Polytechnic State University, San Luis Obispo, California, United States\\
$^{7}$ Central China Normal University, Wuhan, China\\
$^{8}$ Centro de Aplicaciones Tecnol\'{o}gicas y Desarrollo Nuclear (CEADEN), Havana, Cuba\\
$^{9}$ Centro de Investigaci\'{o}n y de Estudios Avanzados (CINVESTAV), Mexico City and M\'{e}rida, Mexico\\
$^{10}$ Chicago State University, Chicago, Illinois, United States\\
$^{11}$ China Institute of Atomic Energy, Beijing, China\\
$^{12}$ Chungbuk National University, Cheongju, Republic of Korea\\
$^{13}$ Comenius University Bratislava, Faculty of Mathematics, Physics and Informatics, Bratislava, Slovakia\\
$^{14}$ COMSATS University Islamabad, Islamabad, Pakistan\\
$^{15}$ Creighton University, Omaha, Nebraska, United States\\
$^{16}$ Department of Physics, Aligarh Muslim University, Aligarh, India\\
$^{17}$ Department of Physics, Pusan National University, Pusan, Republic of Korea\\
$^{18}$ Department of Physics, Sejong University, Seoul, Republic of Korea\\
$^{19}$ Department of Physics, University of California, Berkeley, California, United States\\
$^{20}$ Department of Physics, University of Oslo, Oslo, Norway\\
$^{21}$ Department of Physics and Technology, University of Bergen, Bergen, Norway\\
$^{22}$ Dipartimento di Fisica dell'Universit\`{a} and Sezione INFN, Cagliari, Italy\\
$^{23}$ Dipartimento di Fisica dell'Universit\`{a} and Sezione INFN, Trieste, Italy\\
$^{24}$ Dipartimento di Fisica dell'Universit\`{a} and Sezione INFN, Turin, Italy\\
$^{25}$ Dipartimento di Fisica e Astronomia dell'Universit\`{a} and Sezione INFN, Bologna, Italy\\
$^{26}$ Dipartimento di Fisica e Astronomia dell'Universit\`{a} and Sezione INFN, Catania, Italy\\
$^{27}$ Dipartimento di Fisica e Astronomia dell'Universit\`{a} and Sezione INFN, Padova, Italy\\
$^{28}$ Dipartimento di Fisica e Nucleare e Teorica, Universit\`{a} di Pavia, Pavia, Italy\\
$^{29}$ Dipartimento di Fisica `E.R.~Caianiello' dell'Universit\`{a} and Gruppo Collegato INFN, Salerno, Italy\\
$^{30}$ Dipartimento DISAT del Politecnico and Sezione INFN, Turin, Italy\\
$^{31}$ Dipartimento di Scienze e Innovazione Tecnologica dell'Universit\`{a} del Piemonte Orientale and INFN Sezione di Torino, Alessandria, Italy\\
$^{32}$ Dipartimento di Scienze MIFT, Universit\`{a} di Messina, Messina, Italy\\
$^{33}$ Dipartimento Interateneo di Fisica `M.~Merlin' and Sezione INFN, Bari, Italy\\
$^{34}$ European Organization for Nuclear Research (CERN), Geneva, Switzerland\\
$^{35}$ Faculty of Electrical Engineering, Mechanical Engineering and Naval Architecture, University of Split, Split, Croatia\\
$^{36}$ Faculty of Engineering and Science, Western Norway University of Applied Sciences, Bergen, Norway\\
$^{37}$ Faculty of Nuclear Sciences and Physical Engineering, Czech Technical University in Prague, Prague, Czech Republic\\
$^{38}$ Faculty of Science, P.J.~\v{S}af\'{a}rik University, Ko\v{s}ice, Slovakia\\
$^{39}$ Frankfurt Institute for Advanced Studies, Johann Wolfgang Goethe-Universit\"{a}t Frankfurt, Frankfurt, Germany\\
$^{40}$ Fudan University, Shanghai, China\\
$^{41}$ Gangneung-Wonju National University, Gangneung, Republic of Korea\\
$^{42}$ Gauhati University, Department of Physics, Guwahati, India\\
$^{43}$ Helmholtz-Institut f\"{u}r Strahlen- und Kernphysik, Rheinische Friedrich-Wilhelms-Universit\"{a}t Bonn, Bonn, Germany\\
$^{44}$ Helsinki Institute of Physics (HIP), Helsinki, Finland\\
$^{45}$ High Energy Physics Group,  Universidad Aut\'{o}noma de Puebla, Puebla, Mexico\\
$^{46}$ Hiroshima University, Hiroshima, Japan\\
$^{47}$ Hochschule Worms, Zentrum  f\"{u}r Technologietransfer und Telekommunikation (ZTT), Worms, Germany\\
$^{48}$ Horia Hulubei National Institute of Physics and Nuclear Engineering, Bucharest, Romania\\
$^{49}$ Indian Institute of Technology Bombay (IIT), Mumbai, India\\
$^{50}$ Indian Institute of Technology Indore, Indore, India\\
$^{51}$ Indonesian Institute of Sciences, Jakarta, Indonesia\\
$^{52}$ INFN, Laboratori Nazionali di Frascati, Frascati, Italy\\
$^{53}$ INFN, Sezione di Bari, Bari, Italy\\
$^{54}$ INFN, Sezione di Bologna, Bologna, Italy\\
$^{55}$ INFN, Sezione di Cagliari, Cagliari, Italy\\
$^{56}$ INFN, Sezione di Catania, Catania, Italy\\
$^{57}$ INFN, Sezione di Padova, Padova, Italy\\
$^{58}$ INFN, Sezione di Pavia, Pavia, Italy\\
$^{59}$ INFN, Sezione di Torino, Turin, Italy\\
$^{60}$ INFN, Sezione di Trieste, Trieste, Italy\\
$^{61}$ Inha University, Incheon, Republic of Korea\\
$^{62}$ Institute for Gravitational and Subatomic Physics (GRASP), Utrecht University/Nikhef, Utrecht, Netherlands\\
$^{63}$ Institute for Nuclear Research, Academy of Sciences, Moscow, Russia\\
$^{64}$ Institute of Experimental Physics, Slovak Academy of Sciences, Ko\v{s}ice, Slovakia\\
$^{65}$ Institute of Physics, Homi Bhabha National Institute, Bhubaneswar, India\\
$^{66}$ Institute of Physics of the Czech Academy of Sciences, Prague, Czech Republic\\
$^{67}$ Institute of Space Science (ISS), Bucharest, Romania\\
$^{68}$ Institut f\"{u}r Kernphysik, Johann Wolfgang Goethe-Universit\"{a}t Frankfurt, Frankfurt, Germany\\
$^{69}$ Instituto de Ciencias Nucleares, Universidad Nacional Aut\'{o}noma de M\'{e}xico, Mexico City, Mexico\\
$^{70}$ Instituto de F\'{i}sica, Universidade Federal do Rio Grande do Sul (UFRGS), Porto Alegre, Brazil\\
$^{71}$ Instituto de F\'{\i}sica, Universidad Nacional Aut\'{o}noma de M\'{e}xico, Mexico City, Mexico\\
$^{72}$ iThemba LABS, National Research Foundation, Somerset West, South Africa\\
$^{73}$ Jeonbuk National University, Jeonju, Republic of Korea\\
$^{74}$ Johann-Wolfgang-Goethe Universit\"{a}t Frankfurt Institut f\"{u}r Informatik, Fachbereich Informatik und Mathematik, Frankfurt, Germany\\
$^{75}$ Joint Institute for Nuclear Research (JINR), Dubna, Russia\\
$^{76}$ Korea Institute of Science and Technology Information, Daejeon, Republic of Korea\\
$^{77}$ KTO Karatay University, Konya, Turkey\\
$^{78}$ Laboratoire de Physique des 2 Infinis, Ir\`{e}ne Joliot-Curie, Orsay, France\\
$^{79}$ Laboratoire de Physique Subatomique et de Cosmologie, Universit\'{e} Grenoble-Alpes, CNRS-IN2P3, Grenoble, France\\
$^{80}$ Lawrence Berkeley National Laboratory, Berkeley, California, United States\\
$^{81}$ Lund University Department of Physics, Division of Particle Physics, Lund, Sweden\\
$^{82}$ Moscow Institute for Physics and Technology, Moscow, Russia\\
$^{83}$ Nagasaki Institute of Applied Science, Nagasaki, Japan\\
$^{84}$ Nara Women{'}s University (NWU), Nara, Japan\\
$^{85}$ National and Kapodistrian University of Athens, School of Science, Department of Physics , Athens, Greece\\
$^{86}$ National Centre for Nuclear Research, Warsaw, Poland\\
$^{87}$ National Institute of Science Education and Research, Homi Bhabha National Institute, Jatni, India\\
$^{88}$ National Nuclear Research Center, Baku, Azerbaijan\\
$^{89}$ National Research Centre Kurchatov Institute, Moscow, Russia\\
$^{90}$ Niels Bohr Institute, University of Copenhagen, Copenhagen, Denmark\\
$^{91}$ Nikhef, National institute for subatomic physics, Amsterdam, Netherlands\\
$^{92}$ NRC Kurchatov Institute IHEP, Protvino, Russia\\
$^{93}$ NRC \guillemotleft Kurchatov\guillemotright  Institute - ITEP, Moscow, Russia\\
$^{94}$ NRNU Moscow Engineering Physics Institute, Moscow, Russia\\
$^{95}$ Nuclear Physics Group, STFC Daresbury Laboratory, Daresbury, United Kingdom\\
$^{96}$ Nuclear Physics Institute of the Czech Academy of Sciences, \v{R}e\v{z} u Prahy, Czech Republic\\
$^{97}$ Oak Ridge National Laboratory, Oak Ridge, Tennessee, United States\\
$^{98}$ Ohio State University, Columbus, Ohio, United States\\
$^{99}$ Petersburg Nuclear Physics Institute, Gatchina, Russia\\
$^{100}$ Physics department, Faculty of science, University of Zagreb, Zagreb, Croatia\\
$^{101}$ Physics Department, Panjab University, Chandigarh, India\\
$^{102}$ Physics Department, University of Jammu, Jammu, India\\
$^{103}$ Physics Department, University of Rajasthan, Jaipur, India\\
$^{104}$ Physikalisches Institut, Eberhard-Karls-Universit\"{a}t T\"{u}bingen, T\"{u}bingen, Germany\\
$^{105}$ Physikalisches Institut, Ruprecht-Karls-Universit\"{a}t Heidelberg, Heidelberg, Germany\\
$^{106}$ Physik Department, Technische Universit\"{a}t M\"{u}nchen, Munich, Germany\\
$^{107}$ Politecnico di Bari and Sezione INFN, Bari, Italy\\
$^{108}$ Research Division and ExtreMe Matter Institute EMMI, GSI Helmholtzzentrum f\"ur Schwerionenforschung GmbH, Darmstadt, Germany\\
$^{109}$ Russian Federal Nuclear Center (VNIIEF), Sarov, Russia\\
$^{110}$ Saha Institute of Nuclear Physics, Homi Bhabha National Institute, Kolkata, India\\
$^{111}$ School of Physics and Astronomy, University of Birmingham, Birmingham, United Kingdom\\
$^{112}$ Secci\'{o}n F\'{\i}sica, Departamento de Ciencias, Pontificia Universidad Cat\'{o}lica del Per\'{u}, Lima, Peru\\
$^{113}$ St. Petersburg State University, St. Petersburg, Russia\\
$^{114}$ Stefan Meyer Institut f\"{u}r Subatomare Physik (SMI), Vienna, Austria\\
$^{115}$ SUBATECH, IMT Atlantique, Universit\'{e} de Nantes, CNRS-IN2P3, Nantes, France\\
$^{116}$ Suranaree University of Technology, Nakhon Ratchasima, Thailand\\
$^{117}$ Technical University of Ko\v{s}ice, Ko\v{s}ice, Slovakia\\
$^{118}$ The Henryk Niewodniczanski Institute of Nuclear Physics, Polish Academy of Sciences, Cracow, Poland\\
$^{119}$ The University of Texas at Austin, Austin, Texas, United States\\
$^{120}$ Universidad Aut\'{o}noma de Sinaloa, Culiac\'{a}n, Mexico\\
$^{121}$ Universidade de S\~{a}o Paulo (USP), S\~{a}o Paulo, Brazil\\
$^{122}$ Universidade Estadual de Campinas (UNICAMP), Campinas, Brazil\\
$^{123}$ Universidade Federal do ABC, Santo Andre, Brazil\\
$^{124}$ University of Cape Town, Cape Town, South Africa\\
$^{125}$ University of Houston, Houston, Texas, United States\\
$^{126}$ University of Jyv\"{a}skyl\"{a}, Jyv\"{a}skyl\"{a}, Finland\\
$^{127}$ University of Kansas, Lawrence, Kansas, United States\\
$^{128}$ University of Liverpool, Liverpool, United Kingdom\\
$^{129}$ University of Science and Technology of China, Hefei, China\\
$^{130}$ University of South-Eastern Norway, Tonsberg, Norway\\
$^{131}$ University of Tennessee, Knoxville, Tennessee, United States\\
$^{132}$ University of the Witwatersrand, Johannesburg, South Africa\\
$^{133}$ University of Tokyo, Tokyo, Japan\\
$^{134}$ University of Tsukuba, Tsukuba, Japan\\
$^{135}$ University Politehnica of Bucharest, Bucharest, Romania\\
$^{136}$ Universit\'{e} Clermont Auvergne, CNRS/IN2P3, LPC, Clermont-Ferrand, France\\
$^{137}$ Universit\'{e} de Lyon, CNRS/IN2P3, Institut de Physique des 2 Infinis de Lyon, Lyon, France\\
$^{138}$ Universit\'{e} de Strasbourg, CNRS, IPHC UMR 7178, F-67000 Strasbourg, France, Strasbourg, France\\
$^{139}$ Universit\'{e} Paris-Saclay Centre d'Etudes de Saclay (CEA), IRFU, D\'{e}partment de Physique Nucl\'{e}aire (DPhN), Saclay, France\\
$^{140}$ Universit\`{a} degli Studi di Foggia, Foggia, Italy\\
$^{141}$ Universit\`{a} di Brescia, Brescia, Italy\\
$^{142}$ Variable Energy Cyclotron Centre, Homi Bhabha National Institute, Kolkata, India\\
$^{143}$ Warsaw University of Technology, Warsaw, Poland\\
$^{144}$ Wayne State University, Detroit, Michigan, United States\\
$^{145}$ Westf\"{a}lische Wilhelms-Universit\"{a}t M\"{u}nster, Institut f\"{u}r Kernphysik, M\"{u}nster, Germany\\
$^{146}$ Wigner Research Centre for Physics, Budapest, Hungary\\
$^{147}$ Yale University, New Haven, Connecticut, United States\\
$^{148}$ Yonsei University, Seoul, Republic of Korea\\

\end{flushleft} 
\end{document}